\documentclass[a4paper,fleqn,usenatbib]{mnras}

\usepackage{mathptmx}
\usepackage{multirow}
\usepackage[T1]{fontenc}
\usepackage{ae,aecompl}
\usepackage{threeparttable}

\usepackage{graphicx}	% Including figure files
\usepackage{amsmath}	% Advanced maths commands
\usepackage{amssymb}	% Extra maths symbols
\usepackage{fixltx2e}
\usepackage{hyperref}
\newcommand{\Msol}{\,$M_{\odot}$\,}
\newcommand{\Msolyr}{\,$M_{\odot}$\,yr$^{-1}$\,}
\newcommand{\Rsol}{\,$R_{\odot}$\,}
\newcommand{\yr}{\,yr$^{-1}$\,}
\newcommand{\kms}{\,km\,s$^{-1}$\,}

\title[Linear polarization of magnetic hot stars]{Untangling magnetic massive star properties with linear polarization variability and the Analytic Dynamical Magnetosphere model}

\author[M. S. Munoz et al.]{
	M. S. Munoz,$^{1}$\thanks{E-mail: 16msm5@queensu.ca}
	G. A, Wade,$^{2,1}$
	D. M. Faes,$^{3}$
	A.C. Carciofi$^{4}$
	and	J. Labadie-Bartz$^{4}$
	\\
	$^{1}$Department of Physics, Engineering Physics and Astronomy, Queen's University, 64 Bader Lane, Kingston, K7L 3N6, ON Canada\\
	$^{2}$Department of Physics and Space Science, Royal Military College of Canada, 13 General Crerar Crescent, Kingston, K7K 7B4, ON, Canada\\
	$^{3}$Gemini Observatory/NSF's NOIRLab, 670 N. Aohoku Place, Hilo, HI 96720, USA\\
	$^{4}$Instituto de Astronomia, Universidade de Sao Paulo, Rua do Matao 1226, Sao Paulo 05508-900, Brazil
}

\date{Accepted XXX. Received YYY; in original form ZZZ}

\pubyear{2021}

\begin{document}
	\label{firstpage}
	\pagerange{\pageref{firstpage}--\pageref{lastpage}}
	\maketitle
	
	% Abstract of the paper
	\begin{abstract}
		
{Light scattered off particles can become linearly polarized. Stars surrounded by oblique, co-rotating envelopes are therefore expected to manifest periodic linear polarimetric variations. The electron scattering magnetospheres of magnetic massive stars are expected to be suitable candidates to observe this effect.  In this paper, we present the first semi-analytical model capable of synthesizing the continuum polarimetric signatures of magnetic O-type stars in an optically thin, single electron scattering limit. The purpose of this investigation is to improve our general understanding of magnetic hot stars by characterizing their polarimetric behaviour. Our linear polarization model is constructed by combining the analytical expressions for the polarimetric variations of an obliquely rotating envelope with the Analytic Dynamical Magnetosphere model to represent a physical model for the envelope density structure. We compute grids of model Stokes $Q$ and $U$ curves and show that their shapes are unique to the choice of inclination and obliquity angles. We apply our model to HD 191612, a prototypical Of?p-type star, having both polarimetric and photometric observations.  We find that the polarimetric modulations are best reproduced with $i=19^{+12}_{-3}$\,$^\circ$, $\beta=71^{+3}_{-9}$\,$^\circ$, and $\log \dot{M}_{B=0}=-6.11^{+0.12}_{-0.06}$ [M$_{\odot}$ yr$^{-1}$]. These results agree with previous investigations of this star. By combining both polarimetric and photometric synthesis tools, we simultaneously model the observations thus adding further refinement of the wind and magnetic properties of HD 191612.}

	\end{abstract}
	
	% Select between one and six entries from the list of approved keywords.
	% Don't make up new ones.. 
	
	%%%%%%%%%%%%%%%%%%%%%%%%%%%%%%%%%%%%%%%%%%%%%%%%%%
	
	%%%%%%%%%%%%%%%%% BODY OF PAPER %%%%%%%%%%%%%%%%%%

	\section{Introduction} \label{intro}
	
	Magnetic massive stars are a rare class of objects. In the search for Magnetism in Massive Stars (MiMeS), a survey which comprised more than 100 Galactic O-type stars, a mere $7\%$ of the sample were found to be magnetic \citep{Grunhut2017}. Their rarity has sparked great interest in the massive star community. Indeed, magnetic hot stars present a unique opportunity to investigate the dynamical interactions between their inherently strong stellar winds and magnetic fields that are theoretically predicted and observed to lead to the formation of complex circumstellar, wind-fed magnetospheres. 

	In total, stellar magnetic fields have been firmly detected in only 11 Galactic O-type stars. Their magnetic fields are characteristically strong, stable and organized in a dipolar or low-order multipolar configuration. From an observational standpoint, the magnetic axis has often been found to be inclined with respect to the rotation axis of the star, resulting in the manifestation of rotationally modulated observable quantities \citep[e.g.][]{Wade2011}. In compliance with the paradigm of an oblique rotator model \citep[ORM,][]{ORM}, such magnetospheric signatures are pivotal for diagnosing magnetic hot stars.

	Among the known Galactic magnetic O-type stars, half belong to the peculiar class of Of?p-type stars. Defined by \cite{Walborn1972}, Of?p-type stars are identified by the presence of N {\sc iii} $\lambda$4634 - 41 lines in emission that have comparable strengths to their neighboring C {\sc iii} $\lambda$4650 lines. In order of discovery, there are 6 known Galactic Of?p-type stars \citep[e.g.][]{Walborn2010}: HD 108, HD 191612, HD 148937, NGC 1624-2, CPD-28 2561 and $\theta^1$ Ori C.  They all display periodic variability of numerous observable quantities. Associated with the rotational modulations of an obliquely rotating magnetosphere, their spectroscopic and photometric variability are commonly observed and can be successfully interpreted within the ORM framework \citep[e.g.][]{Townsend2005,Sund2012}. However, rarely is the polarimetric variability observed and even less so compared with theoretical models.
	
		%Polarization by electron scattering
	As unpolarized light passes through a medium it can become partially polarized. In particular, the scattering of star light embedded within gaseous material can produce continuum linear polarization. \citet{BMEI} were first to derive analytical expressions describing the scattered flux produced by an optically thin, electron scattering stellar envelope of arbitrary distribution. This approach has since been adapted by  \citet{Fox1992} to consider obliquely rotating envelopes - also providing semi-analytical formulae for the characterization of their rotationally modulated polarimetric variability. The obliquely rotating magnetospheres that surround massive stars are well-suited to observe and verify this phenomenon. However, this theory has yet to be coupled with a physically motivated model for the density structure of the scattering medium.

	%The magnetospheres that surround massive stars are well-suited to observe and verify this phenomenon. \citet{Fox1992} had adapted this approach for obliquely rotating envelopes and also provided analytical formulae predicting their rotationally modulated polarimetric variability. However, this theory has yet to be coupled with a physically motivated model for the density structure of the scattering medium. 

	The first and currently only detection of a massive star magnetosphere via continuum linear polarization was accomplished by \citet{Carciofi2013}. They attempted to model the polarimetric observations of $\sigma$ Ori E, a magnetic Bp-type star, with the Rigidly Rotating Magnetosphere model \citep[RRM,][]{Townsend2005}. The RRM model is a semi-analytic approach for the magnetosphere modelling of centrifugal magnetospheres that are typically associated with magnetic B-type stars. \citet{Carciofi2013} found success in reproducing the linear polarimetric variability of  $\sigma$ Ori E with a corotating disk and blob model that was physically motivated by the RRM model. 
	
	Recently, an Analytic Dynamical Magnetosphere (ADM) model has been developed by \citet{Owocki2016} that can quickly estimate the large-scale density, velocity and temperature structure of dynamical magnetospheres. Magnetic O-type stars are generally slow rotators that are expected to harbor dynamical magnetospheres. Prior to ADM, most massive star magnetosphere calculations were more formally solved utilizing  sophisticated 2D and 3D magnetohydrodynamic (MHD) simulations by \citet{UDDoula2008,UDDoula2009}. Success in the MHD simulations have been found in reproducing spectral variability of magnetic massive stars \citep[e.g.,][]{Sund2012,UDDoula2013}. In conjunction, the ADM model has been shown to be in good agreement with their MHD counterpart \citep{Owocki2016}, but with the added advantage of being far more computationally efficient. 

	Here, we present the first magnetospheric scattering calculations with the ADM model for polarimetric modeling. For this purpose, we have augmented the ADM model with the polarimetric treatment from \citet{Fox1992} to serve as an ADM-based polarimetric synthesis tool. We exploit our ADM-based polarimetric model as a diagnostic tool to characterize the linear polarimetric variability of magnetic massive stars. Our objective is to determine, constrain or improve important stellar and wind parameters by matching models to observations. This will contribute to better our understanding of the wind and magnetic processes that occur in magnetic massive stars and their immediate environment.
	
	The paper is organized as follows. In section \ref{method}, we will describe the numerical model capable of producing synthetic $Q$ and $U$ curves and then explore the parameter space of the model in section \ref{parameters}. This will be followed by a direct application of our model to HD 191612, a member of the peculiar Of?p class of stars. We conclude in the final section.

	\section{The numerical method} \label{method}
	
	The variable linear polarization produced by an ORM is hypothesised to arise from the periodic change of geometry of an aspherical envelope as it rotates around the star. Indeed, the viewing angle dependent column density causes periodic partial occultations of the star's light by its own envelope envelope.
	
	This effect has been extensively studied by \cite{Fox1992} who provided analytic expressions for both Stokes $Q$ and $U$ linear polarization parameters in the optically thin, single electron scattering limit. These approximations are well-suited for the winds of hot massive stars.
	
	%that are typically optically thin \citep[e.g.][]{Munoz2020} and predominantly governed by the electron scattering opacity \citep[e.g.][]{}.

	\subsection{Linear polarization model}
	
	\subsubsection{Arbitrary envelope}
	Consider a point light source embedded within an arbitrarily shaped, obliquely rotating, electron envelope. We will assume that the envelope rotates uniformly and that the scattering medium is optically thin (i.e. single electron scattering). 
	
	A series of rotation matrices must be applied to the Stokes flux parameters in the observer's frame, in order to obtain them in the oblique (i.e. corotating) frame. Starting from the oblique frame, where the magnetic axis is aligned with an arbitrary $z-$axis, the sequence of transformations is:
	\begin{enumerate}
	    \item rotation by $\beta$ to incline the magnetosphere and align the rotational axis with the $z-$axis, where $\beta$ is the obliquity angle;
	    \item rotation by $\phi$ to rotate the envelope, where $\phi$ is the rotational phase; 
	    \item rotation by $i$ to incline the magnetosphere and align the observer's line-of-sight with the $z-$axis, where $i$ is the inclination angle.
	\end{enumerate}

	After some manipulation, the normalized linear polarization Stokes parameters are given by the following equations, as in \citet[][]{Fox1992}; however, with a misprint corrected\footnote{An error of a factor of two was noted in eqs. 5 and 6 of \citet{Fox1992}. The corrected forms are shown in eqs. \ref{eq:Q1} and \ref{eq:U1} of this paper. In addition, the geometry adopted in \citet{Fox1992} is inconsistent with the geometry that is more commonly adopted in \citet{Munoz2020}. We therefore apply a change of variable to eqs. 5 and 6 of \citet{Fox1992} to be consistent with the work of \citet{Munoz2020}: $\beta \rightarrow -\beta$ and $\phi \rightarrow \phi + \pi/2$.}:
		\begin{equation}\label{eq:Q1}
			\begin{aligned}
			Q = &\frac{1}{2} (\tau_0 - 3\tau_0 \gamma_0)\left[ \sin^2 i (3 \cos^2 \beta - 1) - \sin 2i \sin 2\beta \cos \phi \right.\\
			 &+ \left.(1+\cos^2 i) \sin^2 \beta \cos 2\phi  \right] \\
			&+ \tau_0\gamma_1 \left[ -\sin 2i \cos \beta \sin \phi + (1 + \cos^2 i)\sin \beta \sin 2\phi  \right] \\
			&+ \frac{1}{2}\tau_0\gamma_2 \left[-3 \sin^2i \sin 2\beta + 2 \sin 2i \cos 2\beta \cos \phi \right.\\
			&+ \left. (1 + \cos^2 i)\sin 2\beta \cos 2\phi   \right] \\
			&+ \frac{1}{2}\tau_0\gamma_3 \left[ 3 \sin^2i \sin^2\beta  + \sin 2i \sin 2\beta \cos \phi \right.\\
			&+ \left. (1 + \cos^2 i)(1 + \cos^2 \beta) \cos 2\phi   \right] \\
			&- \tau_0\gamma_4 \left[ -\sin 2i \cos \beta \cos \phi + (1 + \cos^2 i)\cos \beta \sin 2\phi  \right], \\			
			\end{aligned}
		\end{equation}
		\begin{equation}\label{eq:U1}
		\begin{aligned} 
			U=& -(\tau_0 - 3\tau_0 \gamma_0)\left[\sin i \sin 2\beta \sin \phi - \cos i \sin^2 \beta \sin 2\phi \right] \\
			&+\tau_0\gamma_1 \left[ \sin i \cos \beta \cos \phi - \cos i \sin\beta \cos 2\phi \right] \\
			&+\frac{1}{2}\tau_0\gamma_2 \left[-2 \sin i \cos 2\beta \sin \phi + \cos i \sin 2\beta \sin 2\phi    \right] \\
			&+\frac{1}{2}\tau_0\gamma_3 \left[ \sin i \sin 2\beta \sin \phi + \cos i (1+\cos^2\beta) \sin 2\phi   \right]  \\ 
			&+\frac{1}{2}\tau_0\gamma_4 \left[ -\sin i \sin \beta \cos \phi + \cos i \cos\beta \cos 2\phi   \right],
		\end{aligned}
		\end{equation}
	where $\tau_0$ and the $\tau_0\gamma_i$ $(i=0,1,2,3,4)$ terms are weighted integral moments. 
		
	The integral moments describe the density structure of the scattering medium. They are volume integrals that scope the entire electron scattering region. Their explicit forms are \citep[see also][]{Fox1991}  
	\begin{equation}
	    \begin{aligned}
	        \tau_0 &= \frac{\sigma_0}{2}\frac{\alpha_e}{m_p} \int_V D(r) \rho(r,\theta,\varphi) \frac{dV}{r^2}, \\
	        \tau_0\gamma_0 &= \frac{\sigma_0}{2}\frac{\alpha_e}{m_p} \int_V D(r) \rho(r,\theta,\varphi) \cos^2 \theta \frac{dV}{r^2}, \\
	        \tau_0\gamma_1 &= \frac{\sigma_0}{2}\frac{\alpha_e}{m_p} \int_V D(r) \rho(r,\theta,\varphi) \sin 2\theta \cos \phi \frac{dV}{r^2}, \\
	        \tau_0\gamma_2 &= \frac{\sigma_0}{2}\frac{\alpha_e}{m_p} \int_V D(r) \rho(r,\theta,\varphi) \sin 2\theta \sin \phi \frac{dV}{r^2}, \\
	        \tau_0\gamma_3 &= \frac{\sigma_0}{2}\frac{\alpha_e}{m_p} \int_V D(r) \rho(r,\theta,\varphi) \sin^2\theta \cos 2\phi \frac{dV}{r^2}, \\
	        \tau_0\gamma_4 &=
	        \frac{\sigma_0}{2}\frac{\alpha_e}{m_p} \int_V D(r) \rho(r,\theta,\varphi) \sin^2\theta \sin 2\phi \frac{dV}{r^2}, \\
	    \end{aligned}
	\end{equation}
	where $\sigma_0 = 3\sigma_T/16\pi$, $\sigma_T$ is the Thomson cross-section, $\rho$ is the envelope mass density, $m_p$ is the proton mass, $\alpha_e$ is the number of free baryons per electron mass, $dV$ is the volume element in spherical coordinates and $D$ is the finite star depolarization correction factor (see below). The integral moments are typically expressed as a function of the electron number density, $n_e$. However, here we have re-expressed the electron number density as a function of total mass density via $n_e = \alpha_e \rho / m_p$. For a fully ionized environment, appropriate for the winds of hot stars, $\alpha_e = (1+X)/2$ where $X$ is the hydrogen number fraction. 

	The $\gamma_i$ terms describe the distribution of matter within the electron scattering envelope. $\gamma_0$ characterizes the general shape of the envelope and ranges from 0 to 1 for an extremely oblate envelope to an extremely prolate envelope. At the transition from oblate to prolate,  $\gamma_0=1/3$ for a spherical envelope. In addition, the $\gamma_1$ and $\gamma_2$ terms characterize the degree of asymmetry about the equatorial plane, while the $\gamma_3$ and $\gamma_4$ terms characterize the degree of asymmetry within the equatorial plane.  Symmetry can greatly simplify eqs. \ref{eq:Q1} and \ref{eq:U1}. For instance, for an envelope that is spherically symmetric, one can show that $\gamma_0=1/3$, $\gamma_1=\gamma_2=\gamma_3=\gamma_4=0$, yielding $Q=U=0$, as expected.  For an envelope that is axisymmetric, a circumstellar magnetosphere for example, one can show that $\gamma_0 \neq 1/3$,  $\gamma_1=\gamma_2=\gamma_3=\gamma_4=0$, yielding eqs. \ref{eq:Q2} and \ref{eq:U2}.  More complex field topologies can be identified with non-zero $\gamma_i$ terms.

     \subsubsection{Axisymmetric envelope}
     Eqs. \ref{eq:Q1} and \ref{eq:U1} are rather lengthy but can be applied to an arbitrary envelope distribution. In the context of stellar magnetospheres, axial symmetry is often implied. We will now consider a point light source embedded within an axisymmetric envelope. In this case, all $\tau_0\gamma_i$ terms vanish except for $\tau_0\gamma_0$. Eqs. \ref{eq:Q1} and \ref{eq:U1} then simplify to
		\begin{equation}\label{eq:Q2}
			\begin{aligned}
			Q=&\frac{1}{2} (\tau_0 - 3\tau_0 \gamma_0)\left[ \sin^2 i (3 \cos^2 \beta - 1) - \sin 2i \sin 2\beta \cos \phi \right.\\
			 &+ \left.(1+\cos^2 i) \sin^2 \beta \cos 2\phi  \right] ,
			\end{aligned}
		\end{equation}	
		\begin{equation}\label{eq:U2}
			\begin{aligned}	
			U=&-(\tau_0 - 3\tau_0 \gamma_0)\left[\sin i \sin 2\beta \sin \phi - \cos i \sin^2 \beta \sin 2\phi \right] . 
			\end{aligned}
		\end{equation}

	For a point light source, the integral moments are constants (i.e. they do not vary in phase). To consider a star of finite size, we must include the effects of depolarization and occultation. 
		
	Depolarization arises as a finite star correction factor to the point light source results.  The depolarization factor was first derived analytically by \citet{Cassinelli1987} and is given by 
	\begin{equation} \label{eq:D}
		D(r) = \sqrt{1-(R_*/r)^2},   
	\end{equation}
	where $r$ is the radial distance from the center of the star and $R_*$ is the radius of the star. The depolarization factor is implemented within each of the integral moments. $D(r)$ rises from zero near the surface of the star to unity at large distances (i.e. $r \gg R_*$), thus attenuating some of the polarization from the finite star case. As a result, the depolarization factor essentially reduces the net polarization predicted from the point light source case. 
		
    Occultation also occurs as a consequence of a finite sized star. The presence of a finite star will cause part of the electron scattering envelope to be occulted by the star. As such, occulted regions will not contribute to the net polarization. The amount of occultation will depend on the observer's line-of-sight. As a consequence, in the finite star case, the integral moments become phase-dependent and non-zero higher-order $\gamma_i$ terms can occur even among axisymmetric scattering geometries, hence general eqs. \ref{eq:Q1} and \ref{eq:U1} should be employed. Occulted regions of the envelope can be removed by  constraining the boundary conditions in the volume integrals of the integral moments at each rotational phase. However, this can only be implemented at the expense of loosing the simplicity of an analytic model. Contrary to the depolarization effect, there is no analytic correction factor that takes into account the occultation effect.
    
    Finally, when observations are involved, it is important to consider the addition of interstellar polarization, $Q_\text{IS}$ and $U_\text{IS}$. This cannot simply be added to eqs. \ref{eq:Q1} and \ref{eq:U1} because we do not know how the static interstellar polarization to the target is oriented with respect to the intrinsic polarization of the star. We must therefore apply a 2D rotation matrix to the theoretical Stokes $(Q,U)$ vector in order to match the observed $(Q',U')$ vector.

    \subsection{Magnetosphere model}

	The functional relations describing the polarimetric variability produced by an obliquely rotating electron envelope relies on the density structure of the electron scattering medium. To compute this in the context of an oblique magnetic rotator, knowledge of the magnetosphere shape and density is required. For this purpose, we exploit the ADM model developed by \citet{Owocki2016}, an analytical model that is designed to quickly map out the physical characteristics of a dipolar dynamical magnetosphere. 
	
	Slowly rotating hot stars tend to develop dynamical magnetospheres \citep[see][]{Petit2013}. Wind plasma that is located within closed dipole field loops becomes magnetically confined and forms the foundation of a circumstellar, rigidly rotating magnetosphere. \citet{Owocki2016} deconstructs the physical formation of a dynamical magnetosphere in three parts: wind outflow, hot post-shock gas and cooled downflow. Each component is computed independently but are superimposed as they are considered to occur simultaneously. There is no temporal evolution in the physical properties of the ADM magnetosphere. Instead, the ADM model is expected to mimic the time-averaged picture of the more sophisticated full MHD simulations. Preliminary comparisons of the ADM model to MHD simulations suggest that they are in good agreement with each other \citep{Owocki2016}.

	The ADM model receives numerous stellar and magnetic parameters as input: the effective temperature ($T_\text{eff}$), stellar radius ($R_*$), stellar mass ($M_*$), terminal velocity ($v_\infty$), mass-feeding rate ($\dot{M}_{B=0}$), and dipolar magnetic field strength ($B_\text{d}$). These parameters control the size and structure of the magnetosphere.  A 2D slice of the magnetosphere is then generated, typically truncated up the Alfv\'{e}n radius ($R_A$), delimiting the maximal radial extent of wind magnetic confinement.  We rotate this image through 360$^\circ$ to obtain a 3D data cube of the magnetosphere (with axial symmetry implied). The electron density can then be estimated from the total density, under the assumption of a fully ionized wind at solar composition. 
		
	There is an additional non-physical parameter to the ADM model that is known as the smoothing length, $\delta/R_*$. The smoothing length is an ad hoc parameter introduced in the cooled downflow component to avoid singularities near the magnetic equator. Ranging from 0 to 1, a larger smoothing length corresponds to larger spatial dispersion of matter about the equatorial plane (see section \ref{amp} for more details). In the following, we fix $\delta/R_* = 0.1$ to be consistent with other ADM-based modelling works \citep[e.g.,][]{Munoz2020,Erba2021}. 

	We stress that the mass-loss rate encoded in ADM is in fact the wind-feeding rate ($\dot{M}_{B=0}$), and not the actual mass-escaping rate ($\dot{M}$). Due to the presence of an external dipolar field, the confined wind material does not effectively escape the star, thus leading to a significant reduction in the global rate of mass loss. Both parameters are related to each other and are equally important to constrain (see section \ref{disc1}). 
	
	Each of the individual ADM components is constructed under the premise of mass-conservation where their base mass-flux is corrected for the presence of a dipolar field. We performed a numerical exercise to verify that we have not added mass to system when we have co-added the mass density of the individual ADM components. Starting from the base of the wind (i.e. sonic point), we integrated the density through the entire magnetosphere volume. We find that after integrating sufficiently far into the wind, the total mass of the ADM components once added up is comparable to the mass of the normal wind outflow of an unmagnetized star - therefore indicating that mass has been conserved after the implementation of a dipolar magnetic field.

	\subsection{Illustrative results} \label{ADM}
	
	\begin{figure*}
	\hspace*{-1.5cm}\includegraphics[width=21cm]{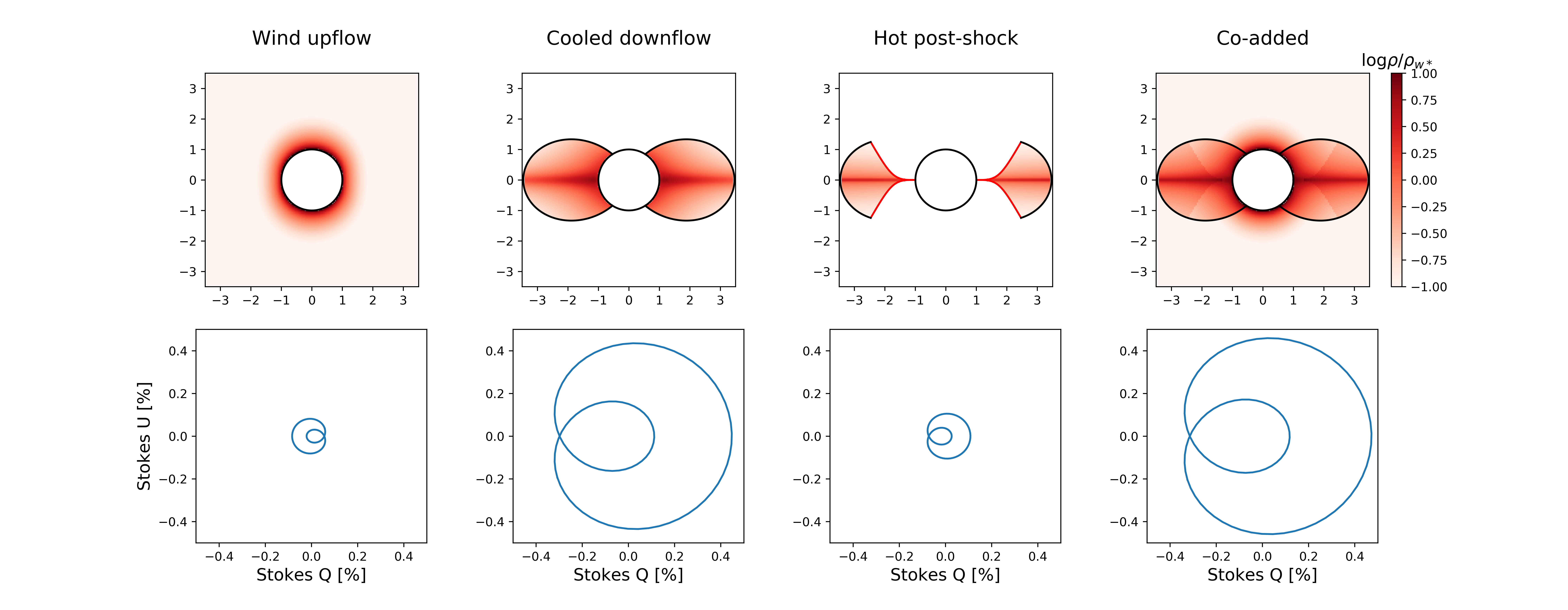}
	\caption{Top: Density structure of the three ADM components (wind upflow, cooled downflow and hot post-shock) and the resulting co-added density. The magnetic axis is horizontal and the densities are normalized to $\rho_{w^*}=\dot{M}_{B=0}/4\pi v_\infty R_*^2$. Bottom: Synthetic Stokes $Q$ and $U$ curves corresponding to the different ADM components and the resulting Stokes $Q$ and $U$ curves from the co-added density.  Linear polarization curves are plotted in $Q-U$ space, tracing the variations along a rotational period.}
	\label{fig:1}
	\end{figure*}
	
	By coupling the ADM formalism with theoretical polarization prescriptions by \citet{Fox1991} (see eqs. \ref{eq:Q2} and \ref{eq:U2}), we can synthesise the linear polarization produced by magnetic hot stars. 
		
	Consider a star of similar stellar and magnetic properties to HD 191612, i.e. $T_\text{eff}=35$\,kK,  $M_*=30$\Msol, $R_*=15$\Rsol, $v_\infty=2700$\kms, $\dot{M}_{B=0}=10^{-6.0}$\Msol \yr and $B_\text{d}=2.5$\,kG  \citep{Wade2012}. With the given physical parameters, the magnetosphere spans $2R_A$ with $R_A = 3.5 R_*$. The top panels of Fig. \ref{fig:1} illustrate the computed ADM density structure of such a star. The three  components of the magnetosphere are plotted with the magnetic axis aligned with an arbitrary $z$-axis. The last (rightmost) panel corresponds to the completed simulated magnetosphere. 
	
	{We can see how the ADM model portrays dipolar wind magnetic confinement. The radiatively driven wind outflow of the underlying star is channeled into the formation of a torus like structure creating density enhancements about the magnetic equator. The work of \citet{Munoz2020} verified that even in the densest part of the magnetosphere, the optically thin, single electron scattering limit still holds (i.e. the electron scattering optical depth remains below unity).}
	%Previous ADM-based electron scattering radiative transfer \citep[see figure 2 of][]{Munoz2020}
	
	For illustration purposes, we now impose the geometric angles and adopt $i=30^\circ$ and $\beta=60^\circ$. The bottom panels of Fig. \ref{fig:1} show the synthesized Stokes $Q$ and $U$ curves, plotted in $Q-U$ space, that are associated to the above density distribution. For simplicity, the polarimetric modulations were computed in the point like source approximation and we neglect interstellar polarization (i.e. $Q_\text{IS}=0$, $U_\text{IS}=0$ and $\Omega=0$). The last (rightmost) panel corresponds to the resulting Stokes $Q$ and $U$ curves computed from the completed magnetosphere. 
	
	The presented Stokes $Q-U$ loci all have a similar shape but different scalings. Analysing eqs. \ref{eq:Q2} and \ref{eq:U2}, we can see that the Stokes $Q$ and $U$ curves share a common amplitude, $\tau_0 - 3\tau_0\gamma_0$. This term solely depends on the density structure of the envelope and serves as a scaling factor to the rotational modulations. The shape of the Stokes $Q-U$ loci are entirely determined by the inclination and obliquity angles. As these angles are fixed, the only variable parameter (between the different lower panels) is the $\tau_0 - 3\tau_0\gamma_0$ amplitude. 

	Looking at the individual Stokes $Q-U$ loci, we notice that the polarimetric variability produced by the wind upflow component is reversed in comparison to the variability produced by the other magnetosphere components. This has to do with distribution of matter within the confined material. The wind upflow component is slightly more prolate, yielding a negative amplitude as $\tau_0 - 3\tau_0\gamma_0 < 0$ (or equivalently $\gamma_0 > 1/3$). The cooled downflow and hot post-shock components are more obviously oblate, yielding a positive amplitude as $\tau_0 - 3\tau_0\gamma_0 > 0$ (or equivalently $\gamma_0 < 1/3$). 
	
	Another important observation is the contribution of the different ADM density components to the overall polarimetric variability. It is apparent that the cooled downflow component constitutes the majority of the magnetosphere density, and as a result, is the major contribution to the linear polarization. The wind upflow component is roughly spherical and therefore does not contribute significantly. For the hot-post shock component, this region is smaller and less dense in comparison to the wind downflow to produce a significant amount of polarization. For completeness, we consider all three components of the ADM model  in our analyses.

	\section{Parameter space study} \label{parameters}
	
	In order to understand the physical implications of the free parameters that enter into a model, it is instructive to perform a parameter space study. The polarimetric model we have just described in Section \ref{method} takes in numerous free parameters. The parameters that describe the magnetosphere are	$T_\text{eff},R_*,M_*, v_\infty$, $\dot{M}_{B=0}$ and $B_\text{d}$. The remaining parameters, $i$ and $\beta$, incline the magnetosphere to form an ORM. 
	
	We will consider a star of fixed stellar parameters but of unknown geometric and magnetic parameters. Among the physical properties that characterise a massive star, the mass-loss rate is the most poorly constrained. For this reason, we will also include the mass-feeding rate in our parameter space study. 
	 
	We will also examine two different cases for treating the central illumination source: a point source, and a light source of finite radius.

	\subsection{Point light source}
	\label{point}
	 
	Illustrated in Figs. \ref{fig:2} and \ref{fig:3} are grids of Stokes $Q$ and $U$ curves plotted in $Q-U$ space.  The $Q-U$  curve tracks the variations of $Q$ and $U$ along an orbital period. Each grid has $i$ and $\beta$ angles varying within the set of $i,\beta=\{10,30,50,70\}$ in degrees. The first grid shows curves of constant $B_\text{d}$ but varying $\dot{M}_{B=0}$. The solid (blue), dashed (orange) and dotted (green) lines correspond to values of $\dot{M}_{B=0}=\{1.0 \times 10^{-6},2.0\times 10^{-6},3.0\times 10^{-6}\}$ in \Msolyr. Similarly, the second grid shows curves of constant $\dot{M}$ but varying $B_\text{d}$. The solid (blue), dashed (orange) and dotted (green) lines correspond to values of $B_\text{d}=\{2.5,5.0,7.5\}$ in kG. The remaining physical parameters are fixed to those of HD 191612 (see section \ref{ADM} for more details).

	\subsubsection{The shape of the $Q-U$ locus}
	
    The Stokes $Q-U$ loci are generally quasi-elliptical. As $\beta \rightarrow 0^\circ$, the loci reduce to a single point (i.e. no variability in the Stokes $Q$ and $U$ curves), while as $\beta \rightarrow 90^\circ$, the loci are neatly elliptical (i.e. sinusoidal variability in the Stokes $Q$ and $U$ curves with unequal amplitude). From lower to higher values of $\beta$, the $Q-U$ loci morph from single- to double-looped and the magnitude of the variability increases. As $i \rightarrow 0^\circ$, the loci are neatly circular (i.e. sinusoidal variability in the Stokes $Q$ and $U$ curves with identical amplitude), while as $i \rightarrow 90^\circ$, the loci fold into a parabola with a central incursion (i.e. double-waved sinusoidal variability in the Stokes $Q$ curve while single-waved for the Stokes $U$ curve). From lower to higher values of $i$, the $Q-U$ loci morph from double to single-looped and the magnitude of the variability increases. In short, for the parameters of HD 191612, the loci are single-looped if $\beta<i$, yet double looped if $\beta>i$. 
 
    The pair of $i$ and $\beta$ angles therefore appear to uniquely determine the shape of the polarimetric variability. This is important for the capability of our method to distinguish the geometric angles via modelling. This is not possible with photometric modelling because the $i$ and $\beta$ angles are degenerate \citep[e.g.][]{Wade2011}. 

    \subsubsection{The amplitude of the $Q-U$ locus}
    \label{amp}
	While the shapes of the Stokes $Q-U$ loci are primarily governed by the geometric angles, the amplitude of the polarimetric variability is specified by the integral moments. The amplitude is maximized when the envelope is either extremely oblate and dense. For a dipolar ORM, this can be accomplished by either increasing the span of the magnetosphere or its density. These two parameters are primarily controlled by the mass-loss rate and dipolar field strength.	On the one hand, the extent of the magnetosphere is determined by the Alfv\'{e}n radius: $R_A \propto B_\text{d}^{1/2} \dot{M}_{B=0}^{-1/4}$ \citep{UDDoula2002}.  On the other hand, the density of the magnetosphere is primarily dependent on the wind density and thus the mass-feeding rate: $\rho \propto \dot{M}_{B=0}$. We can see that an increase of either $\dot{M}_{B=0}$ or $B_\text{d}$ will be tied to an increase in $Q-U$ amplitude. This general trend can be seen in Figs. \ref{fig:2} and \ref{fig:3} respectively. Indeed, we can see that a linear increase of $\dot{M}_{B=0}$ leads to a quasi-linear increase in polarimetric amplitude. However, a linear increase of $B_\text{d}$ leads to a more modest (sub-linear) increase of amplitude. The polarimetric amplitude is therefore far more sensitive to $\dot{M}_{B=0}$ than to $B_\text{d}$. 
	
	{Although not explored here, varying the smoothing length will also affect the amplitude of the polarimetric variability. This effect was more extensively analysed by \citet{Munoz2020} in their ADM-based photometric calculations, also assuming single electron scattering.
	A larger smoothing length would result in weaker polarimetric variability as this would further smooth out the magnetosphere density distribution. Since this quantity can be degenerate with the mass-feeding rate, we chose to fix the smoothing length to a conservative value of 0.1 so as to not overestimate the mass-feeding rate.}

	\subsubsection{The position and orientation of the $Q-U$ locus}
\begin{figure*}
	\includegraphics[width=0.6\linewidth,trim={0 0.25cm 0 1.75cm},clip]{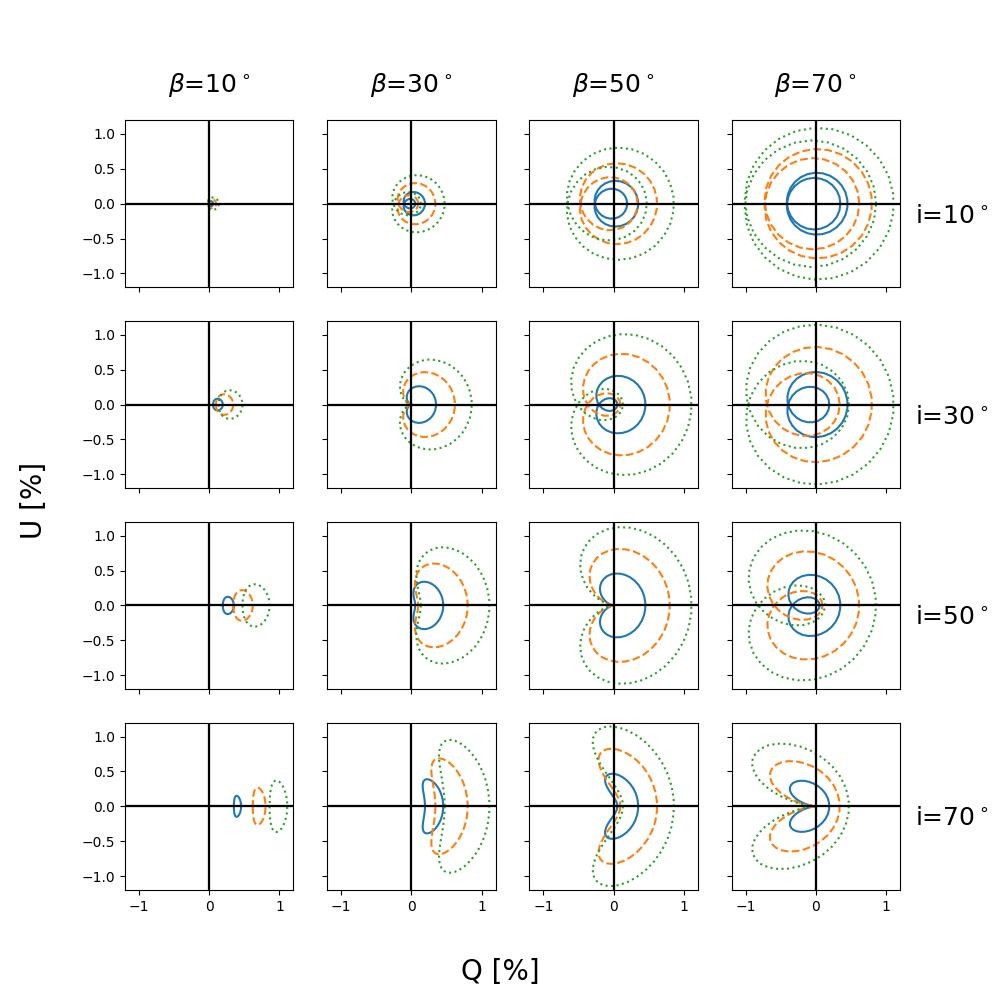}
	\caption{Grid of Stokes $Q$ and $U$ curves computed in the point light source approximation. The fixed parameters correspond to those of HD 191612. Overplotted are curves of varying $\dot{M}_{B=0}$ while $B_\text{d}$ is left constant. The solid (blue), dashed (orange) and dotted (green) lines correspond to values of $\dot{M}_{B=0}=\{1.0,2.0,3.0\}$\Msolyr.}
	\label{fig:2}
	\end{figure*}

\begin{figure*}
	\includegraphics[width=0.6\linewidth,trim={0 0.25cm 0 1.75cm},clip]{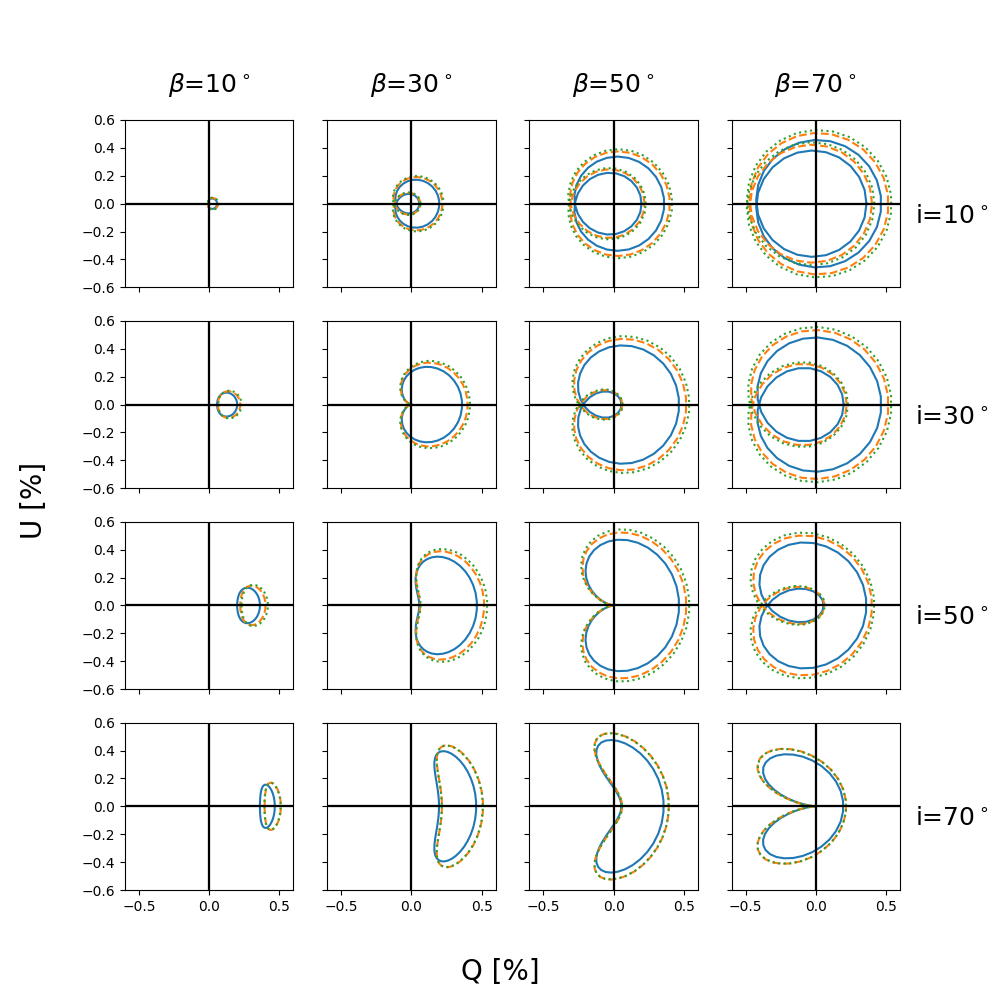}
	\caption{Same as Fig. \ref{fig:2} for varying dipolar field strength, $B_\text{d}$. The solid (blue), dashed (orange) and dotted (green) lines respectively correspond to values of  $B_\text{d}=\{2.5,5.0,7.5\}$\,kG. }
	\label{fig:3}
	\end{figure*}

	The Stokes $Q$ and $U$ curves that are displayed in Figs. \ref{fig:2} and \ref{fig:3} correspond to the intrinsic linear polarization produced by an ORM and are computed in absence of interstellar polarization.  When comparing the intrinsic scattered flux to observations, it is important to incorporate the interstellar polarization, $Q_\text{IS}$ and $U_\text{IS}$, and the rotation angle, $\Omega$, to re-orient the $Q-$axis from the star frame (intrinsic) to the $Q'$-axis of the observer's frame (in general aligned with the North Celestial pole). When applied, interstellar polarization simply displaces the center of the $Q-U$ locus, while the rotation angle, adjusts the orientation of the $Q-U$ locus. Neither of these quantities will affect the shape nor the amplitude of the polarimetric variability. 

	\begin{figure*}
	\includegraphics[width=0.6\linewidth,trim={0 0.25cm 0 1.75cm},clip]{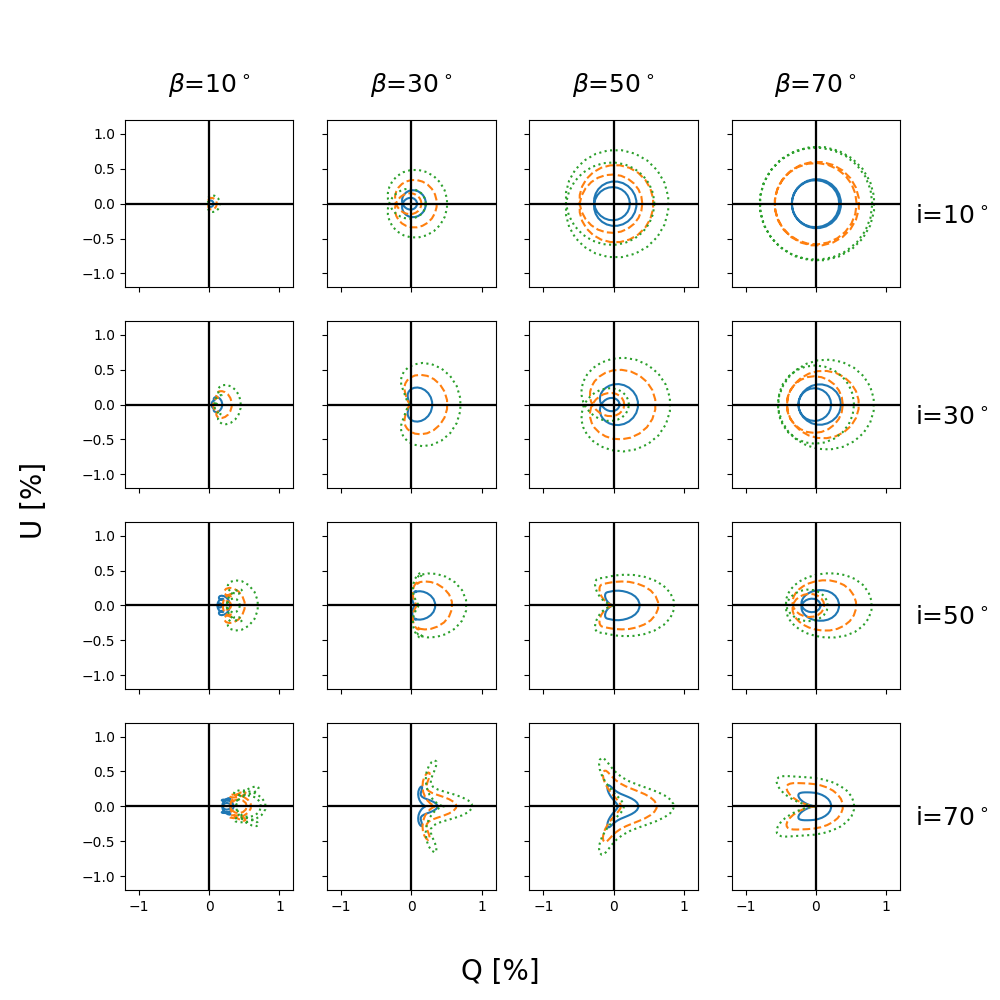}
	\caption{Same as Fig. \ref{fig:2} but for the finite star regime. }
	\label{fig:4}
	\end{figure*}
	
	\begin{figure*}
	\includegraphics[width=0.6\linewidth,trim={0 0.25cm 0 1.75cm},clip]{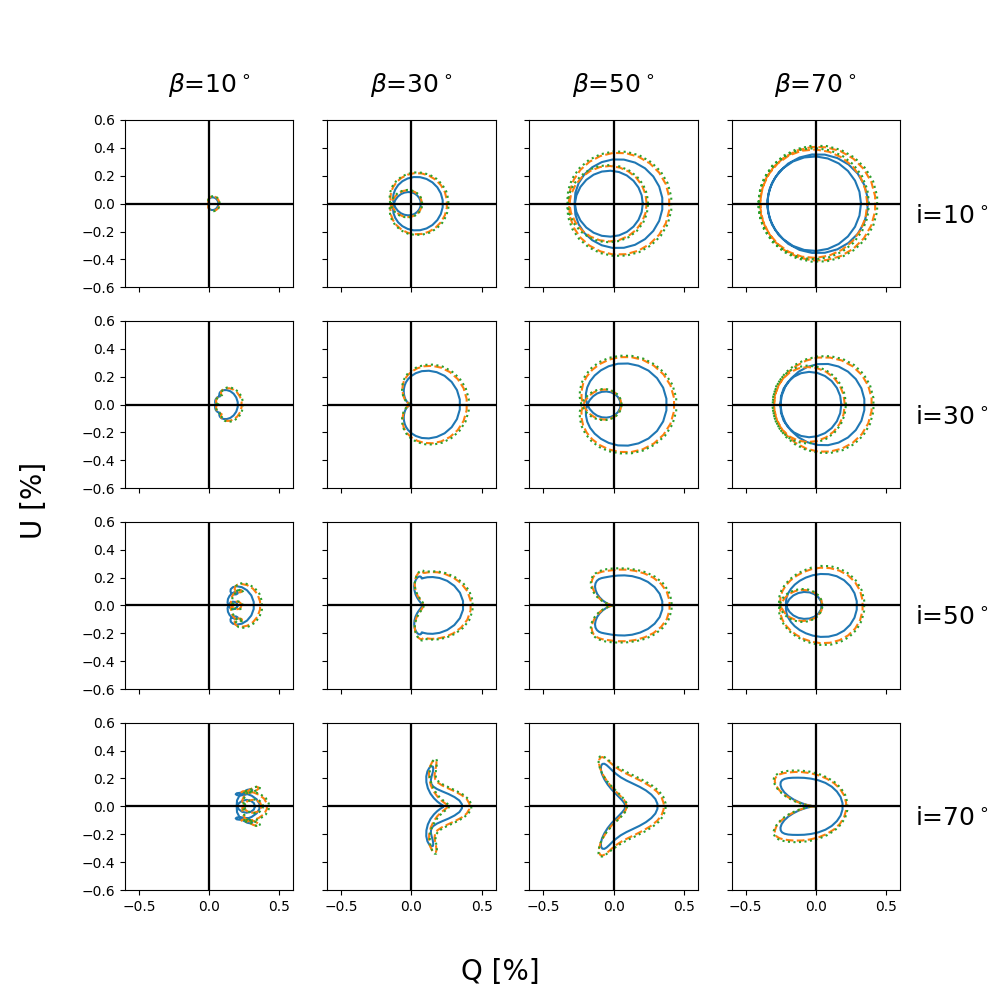}
	\caption{Same as Fig. \ref{fig:3} but for the finite star regime. }
	\label{fig:5}
	\end{figure*}

	\subsection{Finite light source} \label{finite}
 
    Figs. \ref{fig:4} and \ref{fig:5} display grids of Stokes $Q$ and $U$ curves plotted in $Q-U$ space that are calculated in the finite star regime.  The layout of each grid mimics that of the single point light source case described in section \ref{point}. 
	 
    In most cases, the finite star correction leads to a significant reduction in polarization amplitude. This is an expected result on account of the depolarization and occultation effects. However, in certain configurations, particularly at low inclination angles, we note a slight increase in polarization percentage. This can arise from the presence of higher order integral moments that become more important when more material is occulted resulting from global magnetosphere asymmetries that are cause by occultation.
    
    In addition, the shape of the polarization curves in the finite star case are now skewed in comparison to the point light source case. This is a consequence of variable occultation. As the star rotates, sections of the magnetosphere that are behind the star, as seen by the observer, are occulted thus affecting the integral moments or equivalently the amplitude of the polarimetric modulations. Contrary to the point light source case, the amplitude of the polarimetric variability is phase-dependent for a finite-size star. 

    We characterize the amplitude of the polarimetric variability with
		\begin{equation} \label{eq:A}
		    A = \frac{|Q_\text{max} - Q_\text{min}| + |U_\text{max} - U_\text{min}|}{4},
		\end{equation} 
	as defined by \citet{Wolinksi1994}.	To quantify the difference between the point light source results and the finite star results, we compare the ratios of their polarimetric amplitude. We define this ratio as $R = A_\text{f}/A_\text{p}$  where $A_\text{f}$ and $A_\text{p}$ respectively refer to the polarimetric amplitudes of the Stokes $Q$ and $U$ curves computed for a finite star and a point light source. Appendix Table \ref{tab:4} lists the amplitude ratios that were obtained by comparing the curves in Fig. \ref{fig:4} to those in Fig. \ref{fig:2}. Similarly, Table \ref{tab:5} lists the amplitude ratios that were obtained by comparing the curves in Fig. \ref{fig:5} to those in Fig. \ref{fig:3}. The amplitude ratio roughly varies from 1 to 0.5 as function of $B_d$, $i$ and $\beta$.

	\citet{Cassinelli1987} was first to speculate that the point light source approximation can overestimate the polarimetric variability by up to a factor of two. This statement was then further acknowledged by \citet{Fox1992}. Depending on the geometry and structure of the magnetosphere, we confirm that the polarimetric magnitude of the finite star case can be reduced by up to 50\% relative to the point light source case, particularly when the magnetosphere is viewed edge on with respect to the observer's line-of-sight. 

    The polarimetric amplitude ratio can be used to determine correction factors to the point light source model in order to estimate the polarimetric variability in the finite star regime. Figs. \ref{fig:11} and \ref{fig:12} show Stokes $Q$ and $U$ curves that are calculated in the point light source approximation but scaled-down according to the corrections factors listed in Tables \ref{tab:4} and \ref{tab:5}. We can see that at low inclination angles, the $Q-U$ loci computed from the finite star case are in acceptable agreement to the loci computed from the scaled down point light source results. Although the depolarization and occultation effects are not strictly linear, their non-linear effects can be relatively insignificant. Nevertheless, for the rest of the paper, the polarimetric variability will be computed in the finite light source regime (not from the scaled point light source results).

	\section{Application to HD 191612} \label{applications}
 
	HD 191612 is an Of?p-type star that has historically been known for its spectral peculiarities that distinguish it from other Of supergiants \citep{Walborn1973}. Periodic, low-amplitude, spectral variability was first noticed by \citet{Walborn2004}. Coincident magnetic field detections by \citet{Donati2006} lead to the speculation that the variability are of rotational nature.
	
	Extensive monitoring of HD 191612's spectral variability revealed two independent periodicities: a $\sim537$\,d period related to the rotation of the star and a $\sim1542$\,d related to its orbital motion \citep{Howarth2007}. The long term binary period is not expected to significantly affect the short term variability. We therefore assume that the variability of HD 191612 is primarily modulated by the rotational $\sim537$ d period and not by the orbital $\sim1542$\,d period.
	
	HD 191612 was among the first stars to directly confirm the oblique rotator model. \citet{Wade2011} showed that its magnetic, photometric and spectral variations are all phase-locked with respect to the suspected rotational period of $\sim537$\,d. The observed modulations can be interpreted within the framework of an ORM having a dipolar field strength of $\sim2.5$ kG and geometric angles satisfying $i + \beta = 95 \pm 10 ^\circ$. With photometric observations alone it is not possible to distinguish the $i$ and $\beta$ angles. \citet{Wade2011} demonstrated that linear polarimetric observations could help decouple the $i$ and $\beta$ angles; however, observational data was not available at the time. 
	
	In the following subsections we will model the polarimetric variability of HD 191612 as a novel approach to constrain the mass-feeding rate of the star and the magnetic geometry of the magnetosphere. This is particularly valuable for massive stars as their mass-loss rates are difficult to constrain since their actual mass-loss rates are highly sensitive to wind inhomogeneities, such as clumping. We will then revise the photometric modelling with a new data set and finally perform a simultaneous analysis of both the polarimetric and photometric observations.

	\subsection{Fit to the linear polarization}	\label{sec:POLfit}

	\begin{figure*}
	\hspace*{-0.2cm}\includegraphics[width=1.05\linewidth,trim={3.0cm 0 3.0cm 0},clip]{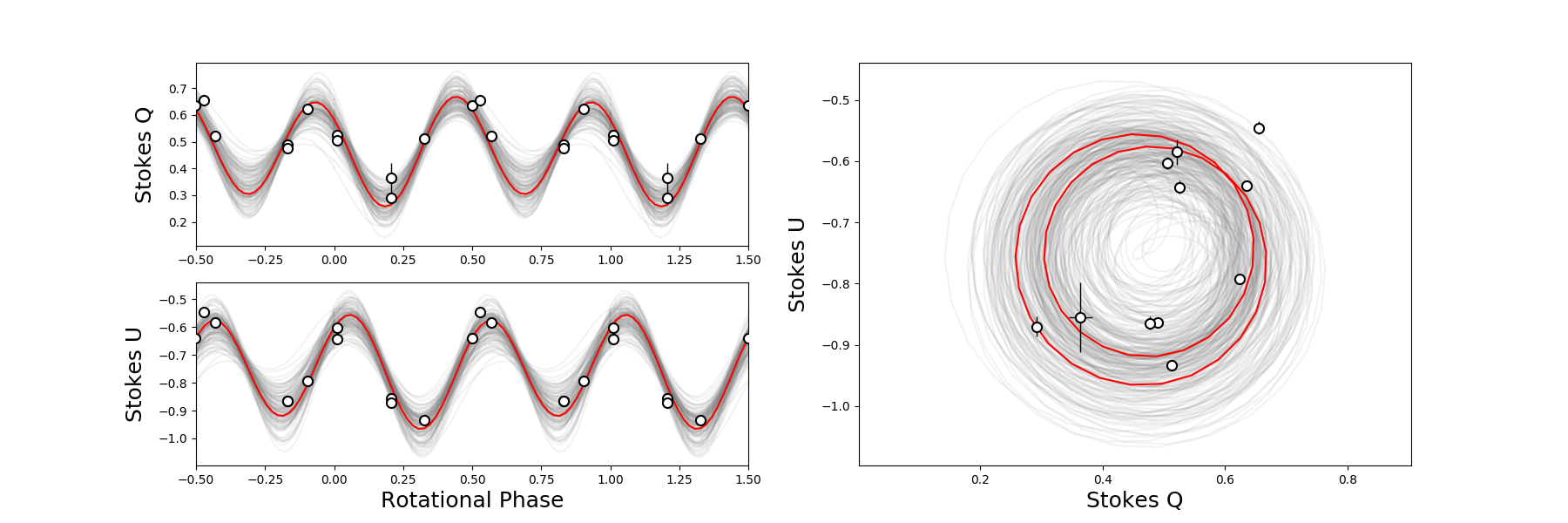}
	\caption{Phased Stokes $Q$ and $U$ parameters for HD 191612. The MCMC-fitted curve is overplotted in red (bold solid curves). Sample curves that span the 1 $\sigma$ error bars on the MCMC-fitted parameters are overplotted in gray (thin solid lines).}
	\label{fig:6}
	\end{figure*}

    The polarimetric observations of HD 191612 were performed in the $V$-band with the IAGPOL polarimeter mounted on the Boller \& Chivens 0.6 m telescope at OPD/LNA, Brazil. Observations and data reduction follow the scheme outlined by \citet{Carciofi2007} and references therein. The phased $Q$ and $U$ curves of HD 191612 are displayed in Fig. \ref{fig:6}. They vary coherently in a sinusoidal-like fashion. We immediately note, based on the results of \ref{parameters}, that both Stokes $Q$ and $U$ curves are double-waved, indicating that $\beta > i$. To model the linear polarization of HD 191612, we utilize the physical parameters that have been derived by \citet{Howarth2007} and \citet{Wade2011}: $T_\text{eff}=35$\,kK, $R_*=14.5$\Rsol, $M_*=30$\Rsol and $v_\infty = 2700$ \kms and $B_\text{d} = 2.5\pm0.4$\,kG. We phase the polarimetric data according to the updated H$\alpha$ ephemeris and period given by \citet{Wade2011}: $\text{JD} = 2 453 415.1(5)\pm537.2(3)\cdot E$.
   	
   We fit eqs. \ref{eq:Q1} and \ref{eq:U1} to the linear polarimetric observations. We carried out the fits using \textsc{emcee}, a Markov-Chain Monte Carlo fitting package \citep{emcee}. Further details on the fitting procedure are provided in section \ref{error}. The MCMC-fitted $Q$ and $U$ curves are illustrated in \ref{fig:6} and the MCMC-fitted parameters are listed in table \ref{tab:1} (see Fig. \ref{fig:13} for a mosaic of the likelihood distributions).
	
   We find that the polarimetric variability is best reproduced with a magnetic dipole model where $i=18^{+10}_{-3}$\,$^\circ$  and $\beta=72^{+3}_{-9}$\,$^\circ$. We obtain a magnetic geometry that is compatible with the results of \citet{Wade2011} where $i$ and $\beta$ were previously constrained to the family of solutions obeying  $i + \beta = 95 \pm 10$\,$^\circ$. In this previous analysis, the $i$ and $\beta$ angles could not be constrained independently as the magnetic geometry was inferred from observational diagnostics (in this case the longitudinal field strength) that rely on the observer's line-of-sight angle. The variation of the magnetic-axis with respect to the observer's line-of-sight is given by: 
\begin{equation} \label{eq2:alpha}
	\cos \alpha = \cos \beta \cos i +  \sin \beta \sin i \cos \phi , 
\end{equation}
   where $\phi$ is the rotational phase. We can see that the $i$ and $\beta$ angles are interchangeable in eq. \ref{eq2:alpha} and are thus degenerate. However, when modelling the linear polarization, the results for the Stokes $Q$ and $U$ curves are unique to $i$ and $\beta$ angles so that they can be determined independently. 

   Here, we have designed the linear polarimetric synthesis tool as mass-feeding rate estimator. We obtain a mass-feeding rate of $\log \dot{M}_{B=0} = -6.14_{-0.12}^{+0.13}$ [\Msol \yr]. This value is comparable to what was previously obtained from spectroscopic modeling. Indeed, \citet{Howarth2007} had reported a clumped mass-loss rate of $\sim  \log \sqrt{f_\text{cl}} \dot{M}_{B=0} = -6.10$ [\Msol \yr] from the H$\alpha$ line where $f_\text{cl}$ is the clumping factor. 
		
   Furthermore, fitting the measured linear polarization curves allows us to estimate the amount of interstellar polarization. We obtain $Q_\text{IS}=0.49_{-0.02}^{+0.02}$\% and $U_\text{IS}=-0.73_{-0.02}^{+0.02}$ \%, corresponding to a polarization percentage of $P_\text{IS}  = 0.88_{-0.02}^{+0.02}$ \%,  with a position angle of $\theta_\text{IS}  = 62_{-9}^{+9}$ $^\circ$. Here the polarization percentage and position angle are respectively defined by 
	\begin{equation}
	    P_\text{IS} = \sqrt{Q_\text{IS}^2 + U_\text{IS}^2},   
	\end{equation}
	and
	\begin{equation}
	    \theta_\text{IS} = 0.5 \arctan{ \left( U_\text{IS}/Q_\text{IS} \right) }.
	\end{equation}
    According to the stellar polarization catalogue compiled by \citet{Heilies2000}, a star within the vicinity of HD 191612 (less than 0.7 degrees in separation) has a polarization percentage of $P_\text{IS}  = 0.550 \pm 0.180$ \% and a position angle of $\theta_\text{IS}  = 64.0 \pm 9 $ $^\circ$. The position angle is in excellent agreement with our results for HD 191612. 

	The linear polarization observations and modeling provide a qualitatively new confirmation of the magnetic oblique rotator model for HD 191612. 

	\begin{table}
	\centering
	\caption{MCMC-fitted parameters to the linear polarimetry of HD 191612}
	\label{tab:1}
	\begin{tabular}{cccccc} % four columns, alignment for each
		\hline
		$i$&$\beta$& $M_\text{B=0}$ & $Q_\text{IS}$ & $U_\text{IS}$ & $\Omega$ \\
		\text{[deg]}&  [deg] &[\Msolyr]& [\%]&[\%] & [deg]\\
		\hline
		$18_{-3}^{+10}$& $72_{-9}^{+3}$&  $-6.14_{-0.12}^{+0.13}$ & $0.49_{-0.02}^{+0.02}$& $-0.73_{-0.02}^{+0.02}$ & $36_{-5}^{+5}$ \\
		\hline
	\end{tabular}
	\end{table}

	\subsection{Fit to the KELT photometry (revised)} \label{sec:LCfit}
	
	Hipparcos observations for HD 191612 were obtained from 1990 to 1993 \citep{HIP}. The phased, folded light curve has been previously analysed and modelled by \citet{Munoz2020}. The recently developed photometric modelling tool is another single electron scattering model that stems from the ADM model, along the same line as the polarimetric modelling tool. In their work, the mass-feeding rate was fixed to constrain the dipole field strength and magnetic geometry. With a mass-feeding rate fixed to $10^{-6.1}$\,\Msol \yr  \footnote{A factor of two was unintentionally omitted in the wind density equation of \citet{Munoz2020} in their implementation of the ADM formalism. While the mass-feeding rate was intended to be fixed at $10^{-5.8}$\,\Msol \yr \citep[e.g.][]{Howarth2007}, the mass-feeding rate was in fact fixed to $10^{-6.1}$\,\Msol \yr with the factor of two in the density taken into consideration.}, they obtained the following MCMC-fitted parameters to the magnetic geometry: $i + \beta=88^{+5}_{-8}$\,$^\circ$, $|i - \beta|= 33^{+26}_{-33}$\,$^\circ$, corresponding to the couple of possible solutions $(i,\beta)=(27_{-14}^{+13}\,^\circ,61_{-11}^{+13}\,^\circ)$, and $B_\text{d}=2.7^{+0.6}_{-0.4}$\,kG. Here, we fix the magnetic field strength and attempt to model the more recently acquired KELT  \citep[Kilodegree Extremely Little Telescope,][]{KELT1,KELT2} light curve as a means to constrain the mass-feeding rate and magnetic geometry. We note that the dipole field strength and mass-fed rate are not typically fit simultaneously as they are degenerate quantities.

    KELT is a ground-based system that consists of two telescopes in the North and South hemispheres. Its primary objective is to search for transiting exoplanets around bright stars. The KELT observations for HD 191612 span from May 30\textsuperscript{th}, 2007 to November 25\textsuperscript{th}, 2014, while the Hipparcos data set spans from November 28\textsuperscript{th} 1989 to February 27\textsuperscript{th}, 1993.  

\begin{figure}
	\hspace{-0.25cm}\includegraphics[width=1.09\columnwidth]{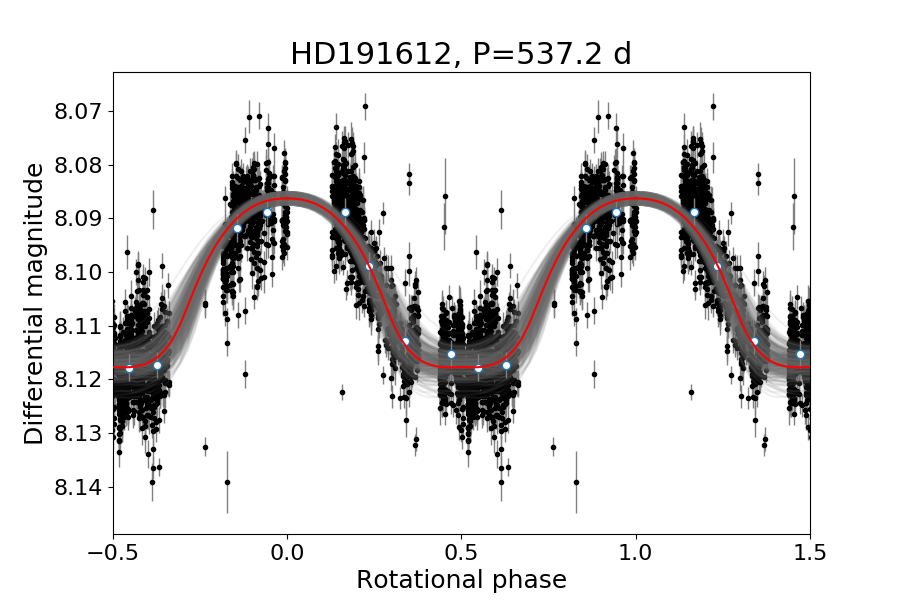}
	\caption{Phased raw (filled circles) and binned (open circles) KELT light curve for HD 191612. The MCMC-fitted curve is overplotted in red (bold solid curves). Sample curves that span the 1 $\sigma$ error bars on the MCMC-fitted parameters are overplotted in gray (thin solid lines).}
	\label{fig:7}
	\end{figure}

    Phasing the KELT light curve according to the ephemeris and period of \citet{Wade2011} reveals a coherent sinusoidal-like curve (see Fig. \ref{fig:7}) that matches the shape and amplitude of the phased Hipparcos light curve \citep[also see Fig. 8 at][]{Munoz2020}. This suggests that the system has remained stable for over two decades. The main advantage of using the  KELT phased light curve is that the inherent dispersion is considerably reduced in comparison to the noise present in the Hipparcos phased light curve. The only drawback of using the ground-base KELT data set, instead of the space-based Hipparcos data set, is that the sampling is irregular, which can lead to significant gaps in the rotational phase coverage.
  
    We adopt a fitting procedure similar to the work of \citet{Munoz2020}. The MCMC-fitted curve to the KELT observations is illustrated in Fig. \ref{fig:7} (see also Fig. \ref{fig:14} for a mosaic of the likelihood distributions). With a magnetic field strength fixed to $B_\text{d} = 2.5$\, kG, we constrain the magnetic geometry and mass-feeding rate to $i+\beta = 89_{-8}^{+5}\,^\circ,\, |i-\beta| = 16_{-11}^{+16}\,^\circ$ - corresponding to the couple of possible solutions $(i,\beta)=(36_{-12}^{+7}\,^\circ,52_{-5}^{+6}\,^\circ)$ - and $\log \dot{M}_{B=0} = -6.02^{+0.13}_{-0.05}$ [\Msol \yr] (see table \ref{tab:2}). These results are not only in agreement with the MCMC-fitted parameters to the Hipparcos photometry, but also appear to be more precise, notably in the $i$ and $\beta$ angles (i.e. decrease in uncertainty range). This is likely due to the reduced scatter in the KELT light curve, however, the $i$ and $\beta$ angles still remain degenerate and cannot be individually identified. 
     	
    \begin{table}
	\centering
	
	\caption{MCMC-fitted parameters to the KELT photometry of HD 191612}
	\label{tab:2}
	
	\begin{threeparttable}
	\begin{tabular}{cccccc} % four columns, alignment for each
		\hline
		$i+\beta$&$|i-\beta|$&  $i$ or $\beta$ & $i$ or $\beta$ & $\dot{M}_\text{B=0}$ & $\Delta m_0^\dagger$\\
		\text{[deg]}&  [deg]  &[deg]&[deg]&[\Msolyr]&[mmag]\\
		\hline
		$89_{-8}^{+5}$& $16_{-11}^{+16}$& $36_{-12}^{+7}$& $52_{-5}^{+6}$ & $-6.02_{-0.05}^{+0.13}$ & $8086_{-1}^{+1}$ \\
		\hline
	\end{tabular}
	
	\begin{tablenotes}
      \small
      \item $^\dagger$ $\Delta m_0$ refers to a constant horizontal offset in the photometric light curve.\hfill
    \end{tablenotes}
    \end{threeparttable}
    \end{table}

	\subsection{Simultaneous fit to the photometry and polarimetry} \label{sec:LCPOLfit}
	
	By simultaneously fitting models to the photometric and polarimetric observations, we may be able to provide additional constraints on the wind and magnetic parameters of HD 191612. For instance, with photometric modelling, it is not possible to distinguish the $i$ and $\beta$ angles from each other, whereas this is not an issue for polarimetric modelling. Even though both observables rely on electron scattering, the photometric variability is primarily sensitive to the line-of-sight optical depth, whereas the polarimetric variability takes into consideration the entire size and shape of the magnetosphere. As a result, the photometric and polarimetric measurements probe different aspects of the massive star magnetosphere that may more sensitively test the predictions of the model and help relieve some degeneracy between the model parameters. 
	
	Our simultaneous fitting of the linear polarization and photometry will include 1) the Hipparcos light curve alone, 2) the KELT light curve alone, and 3) a combination of both the Hipparcos and KELT data. The Hipparcos light curve has regular and complete phase sampling, while the KELT light curve has better precision. Combining the two data sets will therefore be beneficial for improving the phase coverage and signal to noise ratio. 
		
\begin{figure*}
	\includegraphics[width=0.9\linewidth]{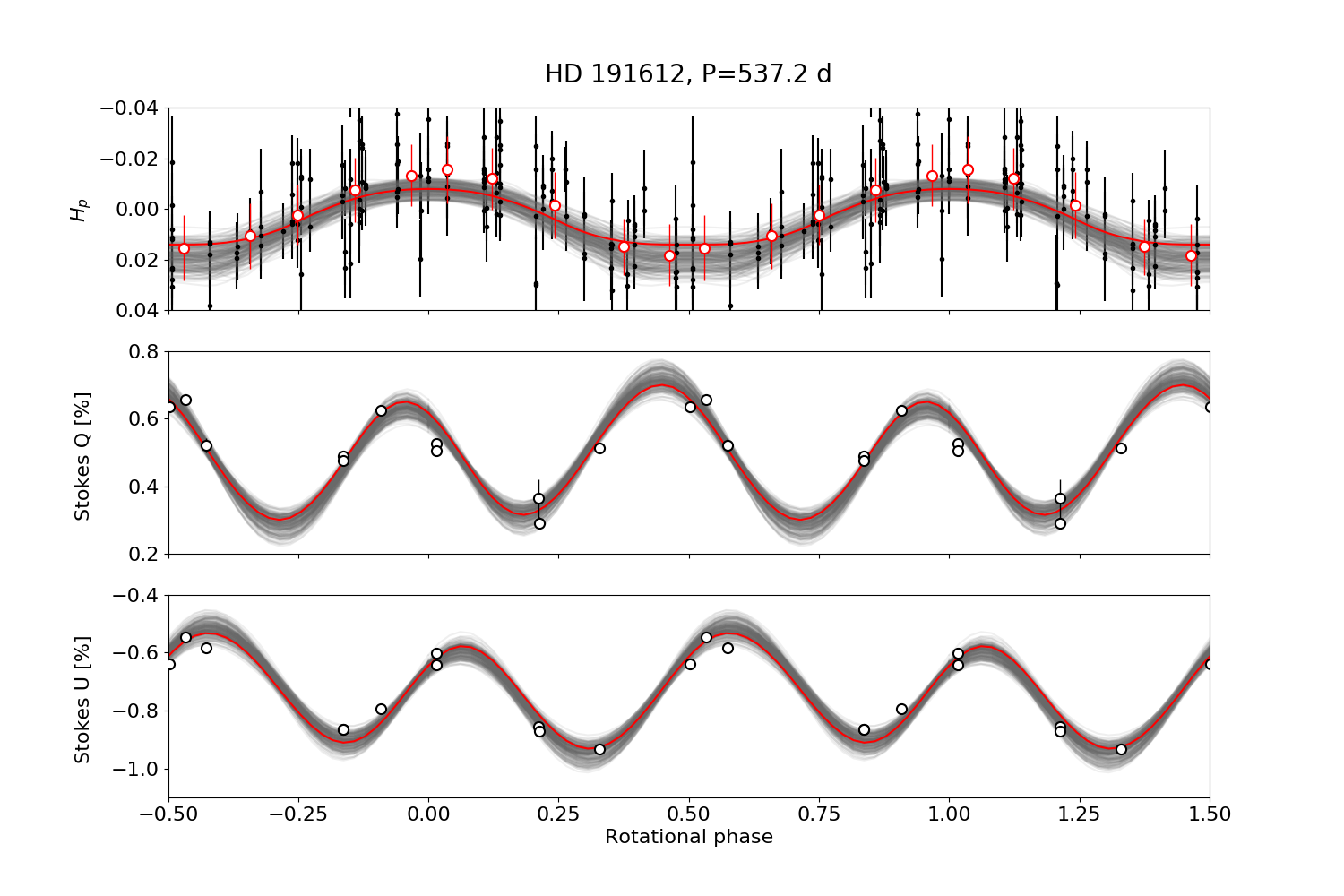}
	\caption{From top to bottom: phased Hipparcos light curve, Stokes $Q$ and $U$ curves. The MCMC-fitted curve, obtained from simultaneously fitting the photometric and polarimetric variability, are overplotted in red (bold solid curves). Sample curves that span the 1 $\sigma$ error bars on the MCMC-fitted parameters are overplotted in gray (thin solid lines).}
	\label{fig:8}
	\end{figure*}

\begin{figure*}
	\includegraphics[width=0.9\linewidth]{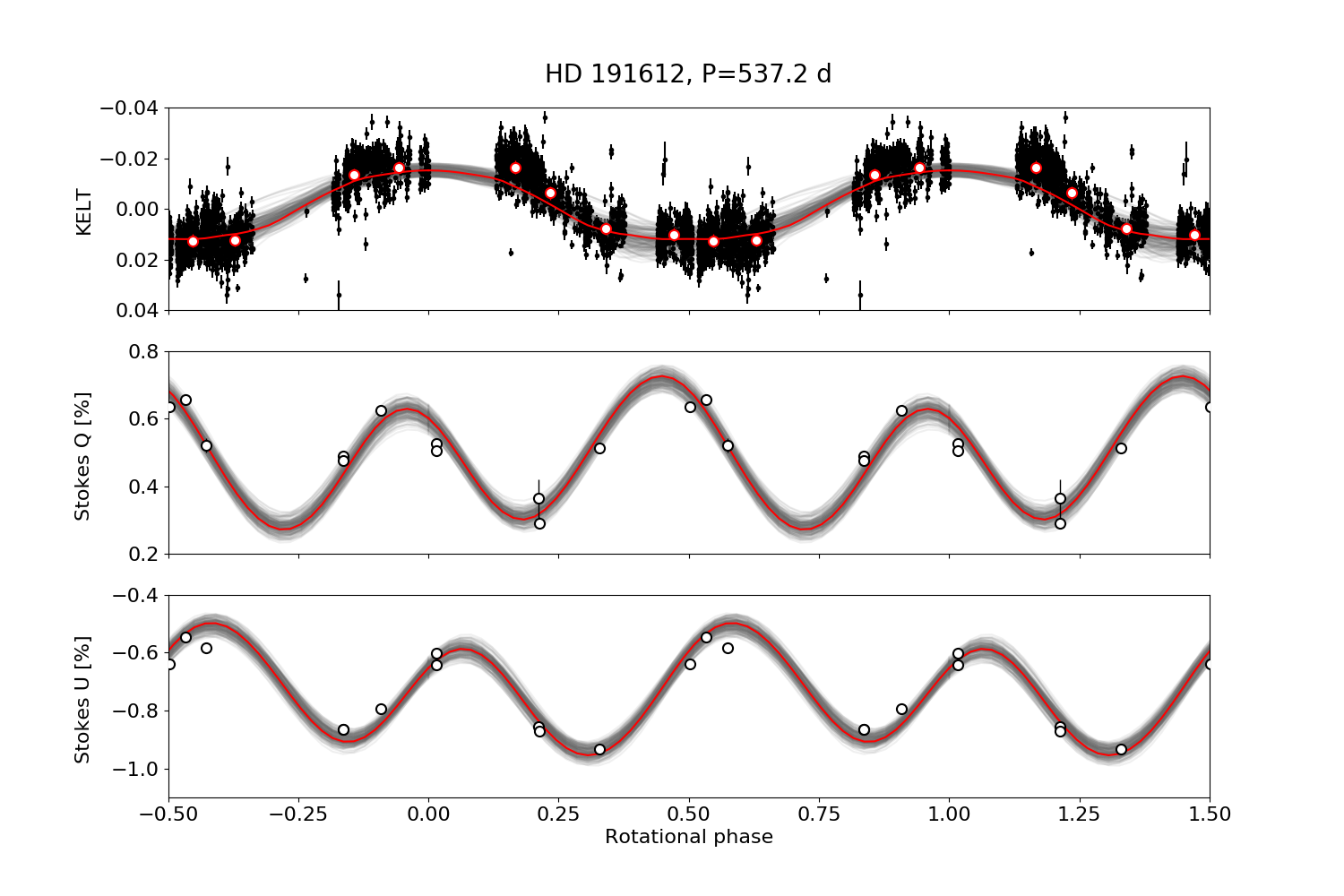}
	\caption{Same as Fig. \ref{fig:8}, but with the KELT light curve.}
	\label{fig:9}
	\end{figure*}

    The results of the simultaneous fits performed are recorded in Table \ref{tab:3}. The MCMC-fitted values are all consistent with each other. The corresponding MCMC-fitted curves are displayed in Figs. \ref{fig:8}, \ref{fig:9} and \ref{fig:11} (see Figs.  \ref{fig:15}, \ref{fig:16} and \ref{fig:17} for a mosaic of the likelihood distributions). It is encouraging to see that a single model can simultaneously reproduce both the photometric and polarimetric variability. The simultaneous fitting efforts did in fact relieve the degeneracy between the $i$ and $\beta$ angles: we were able to rule out the solution with $i>\beta$ derived from the photometric modelling as it is not consistent with the characteristics of the polarimetric variation. 
    
	\begin{figure*}
	\includegraphics[width=0.9\linewidth]{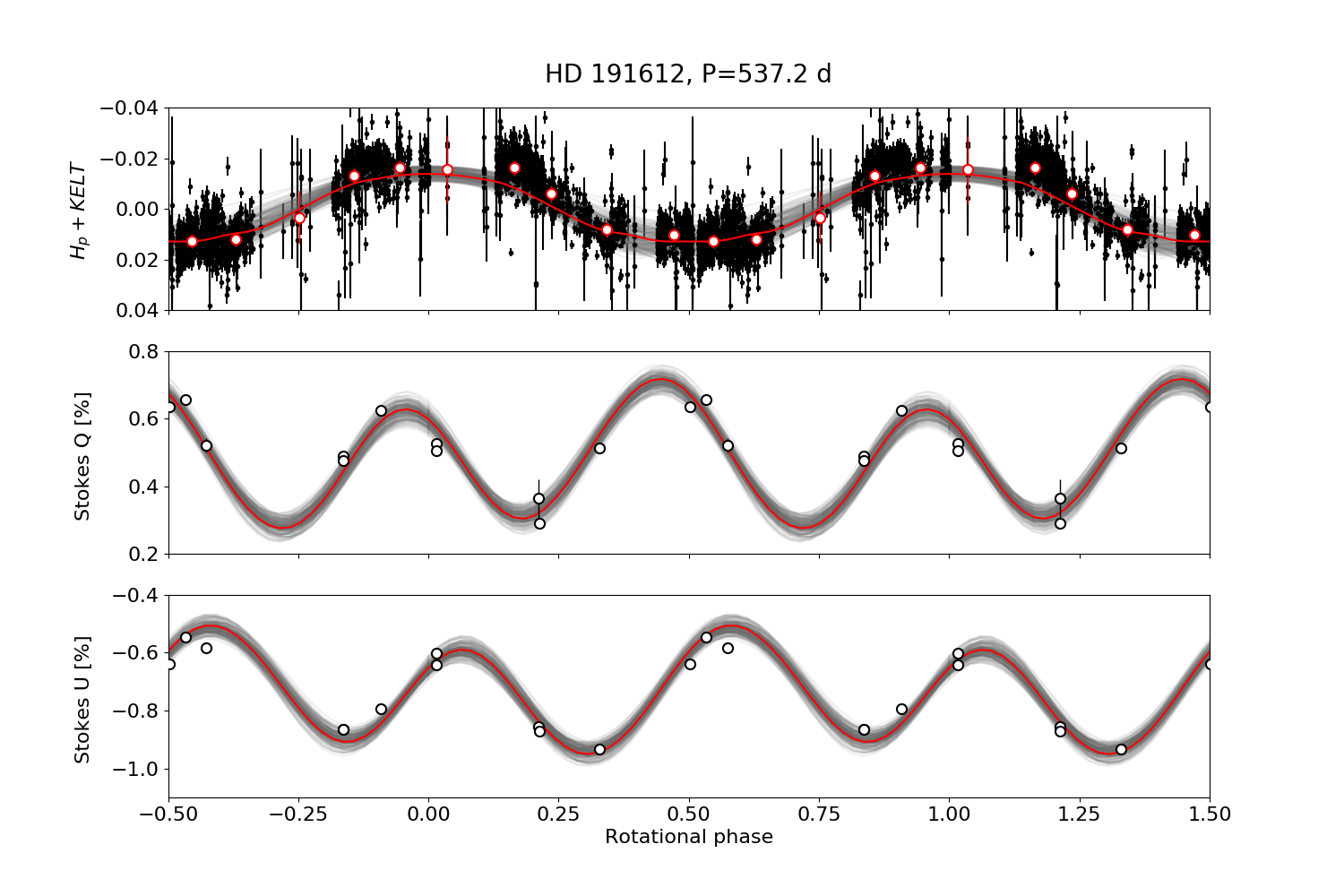}
	\caption{Same as Fig. \ref{fig:8}, but with the combined Hipparcos and KELT light curve. }
	\label{fig:10}
	\end{figure*}

\begin{table*}
	\centering
	\caption{MCMC-fitted parameters to the photometry and polarimetry for HD 191612}
	\label{tab:3}
	\begin{tabular}{cccccccc} % four columns, alignment for each
		\hline
		Photometric data included & $i$ & $\beta$ & $\log \dot{M}$ & $\Delta m_0$ & $Q\textsubscript{IS}$ & $U\textsubscript{IS}$ & $\Omega$  \\ 
		& [deg]  & [deg]  &[\Msolyr]    &   [mmag]     &[\%]         &[\%]                   & [deg]            \\
		\hline
		HIP & $21_{-6}^{+7}$& $68_{-8}^{+6}$ & $-6.14_{-0.06}^{+0.06}$ & $-7.9_{-0.04}^{+0.04}$ & $0.49_{+0.02}^{-0.02}$ & $-0.73_{+0.02}^{-0.02}$ &$36^{+5}_{-5}$ \\
		KELT & $25_{-4}^{+4}$& $66_{-5}^{+4}$ & $-6.10_{-0.06}^{+0.07}$ & $-15.2_{-2.8}^{+2.6}$ & $0.49_{+0.02}^{-0.02}$ & $-0.73_{-0.02}^{+0.03}$ &$35^{+6}_{-6}$ \\
		Both & $24_{-5}^{+5}$& $66_{-5}^{+5}$ & $-6.11_{-0.07}^{+0.06}$ & $-13.8_{-3.1}^{+3.2}$ & $0.49_{+0.02}^{-0.02}$ & $-0.73_{-0.02}^{+0.02}$ &$36^{+6}_{-6}$ \\
		\hline
	\end{tabular}
\end{table*}

	\section{Discussion} \label{discussion}

	\subsection{Orthogonal constraints on the mass-loss rate} \label{disc1}
	%can be more rigorously computed from theoretical line-driven wind models, but is more 
	
    {  The stellar rate of mass loss is an important yet poorly constrained parameter among massive stars. This quantity can be computed from theoretical line-driven wind models, but is more commonly inferred from observational spectral diagnostics. For instance, H$\alpha$ and UV diagnostics are widely used empirical mass-loss rate estimators. }
    
     %As most mass-loss rate determination  their derived mass-loss rates are likely to be overestimated.
    %Most observational mass-loss rate determination tools typically 

    {  Observational mass-loss rate determination tools are generally grouped in two categories: $\rho$-dependent processes and $\rho^2$-dependent processes. $\rho^2$-dependent diagnostics (e.g. H$\alpha$ recombination line) are more sensitive to wind inhomogeneities than $\rho$-dependent diagnostics (e.g. UV resonance lines). As most mass-loss rate diagnostics operate under the assumption of a smooth wind, thus ignoring the presence of clumped density enhancements, $\rho^2$ diagnostics have been found to systematically yield larger mass-loss rates than  $\rho$ diagnostics \citep{Fullerton2006}. This led to the historical introduction of the clumping factor, $f_\text{cl}$, reconciling discordant $\rho$ and $\rho^2$ mass-loss rate estimates. The clumping factor is in fact a measure of small-scale wind inhomogeneities or clumps that effectively lead to a decrease in the inferred rate of mass loss for $\rho^2$ diagnostics, such that $\dot{M}_\text{unclumped} = f_\text{cl}\dot{M}_\text{clumped}$. }
    %Discordant observational mass-loss rate determinations can often be reconciled by introducing a clumping factor, $f_\text{cl}$. 

    { For HD 191612, early H$\alpha$ diagnostics performed by \citet{Howarth2007} returned a mass-loss rate of $1.6 \times 10^{-6}$ \Msolyr, while UV diagnostics by \citet{Marcolino2013} yielded $1.3 \times 10^{-8}$ \Msolyr - more than two orders of magnitude lower than the optical result. We stress that these early reported values were inferred from non-magnetic models. Indeed, they respectively relied on \textsc{FASTWIND}  \citep{Puls2005} and \textsc{CMFGEN} \citep{Hillier1998} models to perform the radiative transfer - both of which assume a spherically symmetric wind, completely ignoring the magnetically confined wind structure. Although the use of non-magnetic models is likely a key cause of the above-mentioned discrepancy in mass-loss rate, \citet{Marcolino2012} stated that it is important to consider the radial extent of the spectral line-formation regions. In fact, UV lines are formed significantly farther out in the wind than optical lines, and likely encompass the Alfv\'{e}n radius. As a result, UV line diagnostics may in fact be probing the rate of mass loss farther out in the wind where the wind is significantly quenched due to the presence of the magnetic field. }
    
   {  HD 191612's observable quantities have recently been revisited using more appropriate magnetic models. In fact, an ADM-based H$\alpha$ line synthesis algorithm has recently been applied to HD 191612 \citep{Owocki2016,Driessen2019}. Preliminary results by \citet{Driessen2019} were able to reproduce the observed EW curve with a clumped mass-feeding rate of $1.1 \times 10^{-6}$ \Msolyr. They emphasize that the absolute mass-loss rate is expected to be much lower once corrected for wind clumping and magnetic wind-trapping. In parallel, an ADM-based UV line synthesis codes have been developed by \citet{Hennicker2018} and \citet{Erba2021}. Their findings confirm that spherically symmetric UV line models cannot be used to accurately derive the wind parameters of magnetic hot stars, thus highlighting the importance of considering non-spherical wind outflows.  A mass-feeding rate of $1.1 \times 10^{-6}$ \Msolyr was adopted by \citet{Hennicker2018} to match theoretical UV resonance-line profiles to the observations. } %It would therefore be of great interest to revise HD 191612's UV line modelling with such an ADM-based UV line synthesis tool. 
 
    { Returning to the ADM-based polarimetric tool described in this paper and the previously developed ADM-based photometric tool \citep{Munoz2020}, they are both $\rho$-dependent models and are therefore insensitive to clumping. When applied independently on HD 191612, we obtained rather large confidence intervals and distorted likelihood distributions (see Fig. \ref{fig:13} and \ref{fig:14}). By combining the tools into a single model, some degeneracy is lifted which further constrains the model input parameters. This explains why the uncertainties in the model parameters are decreased in the simultaneous fits and why the likelihood distributions are rather Gaussian-like, including the obliquity and inclination angles  (see Fig. \ref{fig:17}). It is even more reassuring to see that both electron scattering observational diagnostics are compatible with a common mass-feeding rate of $\log\dot{M}_\text{B=0}=-6.11_{-0.07}^{+0.06}$ [\Msolyr]. In addition, using the \citet{Vink2001} recipe, we obtain a theoretical mass-feeding rate of $10^{-6.1}$ \Msolyr for HD 191612 that is also in agreement with our modelling results. }

    %Both ADM-based polarimetric () and the photometric (previously developped by Munox 2020)  tools are
    %Returning to the 
  
    { The mass-feeding rates constrained above do not represent the effective mass-escaping rate. According to \citet{UDDoula2008}, the dipolar escaping wind fraction (for a non-rotating star) corresponds to 
\begin{equation}
	f_\text{B} = 1 - \sqrt{1-\frac{R_*}{R_\text{c}}},
\end{equation}
    where $R_\text{c}$ is the radius of the last closed loop. The overall magnetic reduction in mass-loss rate is then characterised by $f_B$ where $\dot{M}=f_\text{B} \dot{M}_\text{B=0}$.  Adopting $R_\text{c}\sim R_A$, an Alfv\'{e}n radius of $\sim 3.5\,R_*$ corresponds to a wind quenching of $\sim0.15$. The quenched mass-lass rate for HD 191612 would therefore be reduced by almost an order of magnitude, to $10^{-6.9}$ \Msolyr.   }

    %The UV mass-loss rate from \citet{Marcolino2012} was derived utilizing \textsc{CMFGEN} \citep{Hillier1998} models. \textsc{CMFGEN} assumes a spherically symmetric wind, completely ignoring the magnetically confined wind structure. 
    
    % However, the introduction of a clumping factor does not aid in reconciling our polarimetric mass-feeding rate with the UV results as both methods are  $\rho$-dependent diagnostics. Another mechanism is likely responsible for these discordant measurements. 

    %For our mass-fed rate to be concordant with the determinations of \citet{Howarth2007}, a clumping factor of $f_\text{cl} \sim 4$ is required. 

 %This value is consistent with the value that was derived by \citet{Krti2016}, which is somewhat of a midway point between the UV and H$\alpha$ determinations.

    \subsection{Error assessment} \label{error}
    
    We achieved acceptable fits to the polarimetric variability of HD 191612 within the uncertainty of the available data. Our fits were obtained via MCMC methods, where 100 walkers were initialized, each accomplishing over 1000 steps after disregarding 500 burn-in steps. We adopted a standard Gaussian posterior function where the variance (i.e. the error in the data) is underestimated by some fractional amount, $f$.  The MCMC-fitted model parameters were determined from the peaks likelihood distributions and the confidence intervals were estimated at the 68\% levels. 

    \citet{Wolinksi1994} investigated the confidence intervals of polarimetrically determined inclination angles among binary systems. They noticed a statistical bias towards higher inclination angles. In other words, inclination angles that are derived via polarimetric observations are more likely to be overestimated if the data suffered from large uncertainties. This is an expected result: as the noise of the $Q$ and $U$ curves increases, a straight line becomes a more acceptable fit corresponding to $i\rightarrow 90^\circ$. To counteract this statistical bias, higher signal-to-noise data is therefore required. 

    An analogous statistical bias will thus be present among magnetic oblique rotators. However, in this case, the statistical bias will affect both the magnetic obliquity and the inclination. As $\beta\rightarrow0^\circ$, the Stokes $Q$ and $U$ curves tend to a straight line, regardless of the inclination angle. As $i\rightarrow0^\circ$, the amplitude of the parametric variability is minimal. Thus in general, noisy data will have a statistical bias towards lower obliquity and inclination values \citep[see also][]{Fox1992}.  

    For HD 191612, the quality of the fit mostly suffered from a lack of data points, not from a lack precision. In fact, the double-waved Stokes $Q$ and $U$ curves were sampled with less than a dozen data points. More polarimetric observations obtained at different epochs would therefore be beneficial to the diagnostic precision.

	\subsection{Rotation vs binarity}
	
	The polarimetric variability of an obliquely rotating envelope can be functionally expressed as a $2^\text{nd}$ order Fourier sum in the point light source approximation. By rearranging the terms in eqs.  \ref{eq:Q1} and \ref{eq:U1}, the Stokes $Q$ and $U$ parameters rewrite to
	\begin{equation} \label{eq:Q3}
	Q = q_0 + q_1 \cos(\lambda) + q_2 \sin (\lambda) + q_3 \cos(2 \lambda) + q_4 \sin(2\lambda),
	\end{equation}
	and
	\begin{equation} \label{eq:U3}
	U = u_0 + u_1 \cos(\lambda) + u_2 \sin (\lambda) + u_3 \cos(2 \lambda) + u_4 \sin(2\lambda),
	\end{equation}
	where the $q_i$ and $u_i$ ($i=0,1,2,3,4$) terms are Fourier coefficients. %Analytical expressions for the Fourier coefficients are given in annex for both an arbitrary and axisymmetric envelope. 
	
	Prior to \cite{Fox1992}, linear polarimetric modulations were often thought to be indicative of binary motion. Indeed, \citet{BMEII} had already derived general expressions characterizing polarisation scattering among binary systems, in particular, WR+O binaries \citep[e.g.][]{Drissen1986,StLouis1988} and X-ray binaries \citep[e.g.][]{Rudy1978,Dolan1992}. Such variability can also be expressed in the form of a truncated first and second order Fourier series, equivalent to eqs. \ref{eq:Q3} and eqs. \ref{eq:U3}.
	
	\citet{Fox1992} pointed out the similarities between their own expressions for ORMs and those of \citet{BMEII} for binary stars. It turns out that binary motion can often be misinterpreted as rotational motion or vice versa, at least to a first-order approximation. Though the functional forms are indeed analogous, they differ by their physical meaning, notably in the interpretation of the $q_i$ and $u_i$ terms. However, we note that in the finite star regime, the secondary effects produced by occultations can produce higher order harmonic variations and asymmetries in the $Q-U$ loci that are unique to rotational motion and thus distinguishable from orbital motion. 

	By simultaneously fitting eqs. \ref{eq:Q3} and \ref{eq:U3} to observations, we can determine the Fourier coefficients that best describe the observed linear polarization for HD 191612. This is known as the Fourier coefficient method and has been commonly applied in past studies for binary systems where the fits are typically performed via a minimum least-squares method such as a Levenberg-Marquardt algorithm. 
	
	The advantage of the Fourier coefficient method lies in its simplicity to obtain a curve of best-fit. We can obtain a seemingly adequate simultaneous fit to the Stokes $Q$ and $U$ curves for HD 191612. However, a major drawback is the loss of a physically motivated model. This is because the $q_i$ and $u_i$ terms are fit as independent parameters, when in reality, the Fourier coefficients are all interdependent on the geometric angles and the integral moments. This can lead to fits that are nonphysical in the framework of an ORM (or binary system). 
	In fact, for HD 191612, the best-fit $q_i$ and $u_i$ terms does not represent a physical system when trying to solve for $i$, $\beta$ and $\Omega$. The fitting routine that we have adopted here  (see section \ref{applications} for details) is therefore better suited for extracting the physical model parameters for an obliquely rotating magnetosphere. 
	
	In a generic system, spectroscopic observations are therefore necessary to rule out binary-induced variability. For HD 191612, the rotational period is well constrained via spectroscopy such that we are confident that the polarimetric variability is indeed related to the ORM \citep[see][]{Howarth2007}.

	\section{Conclusion} \label{conclu}

	A light source embedded within an obliquely rotating envelope can give rise to phase-dependent linear polarimetric variability. In this paper, we have developed a simple semi-analytical model designated to estimate the observable linear polarization of  magnetic hot stars. 

	We assume that the electron scattering opacity is responsible for the major contribution of linear polarization - an appropriate assumption for most hot stars.  In this context, we utilize the analytical Stokes $Q$ and $U$ prescriptions of \citet{Fox1992},  derived under the assumption of an optically-thin, Thomson scattering envelope irradiated by a central source. To approximate the linear polarization produced from an obliquely rotating magnetosphere, we exploit the analytic dynamical magnetosphere model by \citet{Owocki2016} to describe the scatterer's distribution.

	We applied our ADM-based polarimetric tool to reproduce the polarimetric signatures of the Galactic Of?p-type star, HD 191612. We were able to constrain its mass-feeding rate and magnetic geometry. By coupling this tool with a previously developed ADM-based photometric tool \citep[see][]{Munoz2020}, we obtained orthogonal constraints on the fundamental wind and magnetic parameters of HD 191612. Our results are consistent with previous findings, but with increased precision. Polarimetric modelling confirms the expected wind density and magnetic confinement from the theory, and therefore is a useful complement to the photometric modelling of this class of stars.

	We acknowledge that we have interpreted the linear polarimetric variations of a HD 191612  within a first-order approximation, i.e single electron scattering. A more accurate depiction would entail a more sophisticated radiative transfer algorithm, including multiple scatterings, NLTE effects and other sources of opacity. Future work will therefore involve comparing our simplistic approach to more sophisticated modelling. 

	Although we have shown just one direct application that benefited from linear polarimetric modelling, we have showcased the ADM model as an attractive alternative method for magnetospheric modelling and have opened the gateway for numerous other applications to stem from this work. Unfortunately magnetic massive stars possessing broadband, linear polarimetric observations are limited. Obtaining more observational data would be essential to further explore the capabilities of the ADM model and the linear polarimetric add-on.

	\section*{Acknowledgements}
The authors acknowledge the careful and thoughtful review by an anonymous referee. G.A.W. acknowledges Discovery Grant support from the Natural Sciences and Engineering Research Council (NSERC) of Canada.	A. C. C. acknowledges support from CNPq (grant 311446/2019-1) and FAPESP (grants 2018/04055-8 and 2019/13354-1). The work of DMF is supported by NOIRLab, which is managed by the Association of Universities for Research in Astronomy (AURA) under a cooperative agreement with the National Science Foundation. J.L.-B. acknowledges support from FAPESP (grant 2017/23731-1).  This work is based on observations made at the Observat\'orio do Pico dos Dias/LNA (Brazil). This project makes use of data from the KELT survey, including support from The Ohio State University, Vanderbilt University, and Lehigh University, along with the KELT follow-up collaboration. This research has made use of NASA's Astrophysics Data System. This research has made use of the SIMBAD database, operated at CDS, Strasbourg, France.

	\section*{DATA AVAILABILITY STATEMENT}
	The codes for the ADM-based light curve and polarimetric synthesis tools  are publicly available at https://github.com/Melissa-Munoz. The data included in this work are available from the authors upon request.

	%%%%%%%%%%%%%%%%%%%%%%%%%%%%%%%%%%%%%%%%%%%%%%%%%%
	
	%%%%%%%%%%%%%%%%%%%% REFERENCES %%%%%%%%%%%%%%%%%%
	
	% The best way to enter references is to use BibTeX:
	
	\bibliographystyle{mnras}
	\bibliography{MSM} % if your bibtex file is called example.bib

	% Alternatively you could enter them by hand, like this:
	% This method is tedious and prone to error if you have lots of references
	%\begin{thebibliography}{99}
	%\bibitem[\protect\citeauthoryear{Author}{2012}]{Author2012}
	%Author A.~N., 2013, Journal of Improbable Astronomy, 1, 1
	%\bibitem[\protect\citeauthoryear{Others}{2013}]{Others2013}
	%Others S., 2012, Journal of Interesting Stuff, 17, 198
	%\end{thebibliography}
	
	%%%%%%%%%%%%%%%%%%%%%%%%%%%%%%%%%%%%%%%%%%%%%%%%%%
	
	%%%%%%%%%%%%%%%%% APPENDICES %%%%%%%%%%%%%%%%%%%%%
	
	\appendix

	\section{Corrections  to the point light source approximation}
	
  		\begin{figure*}
	\includegraphics[width=0.6\linewidth,trim={0 0.25cm 0 1.75cm},clip]{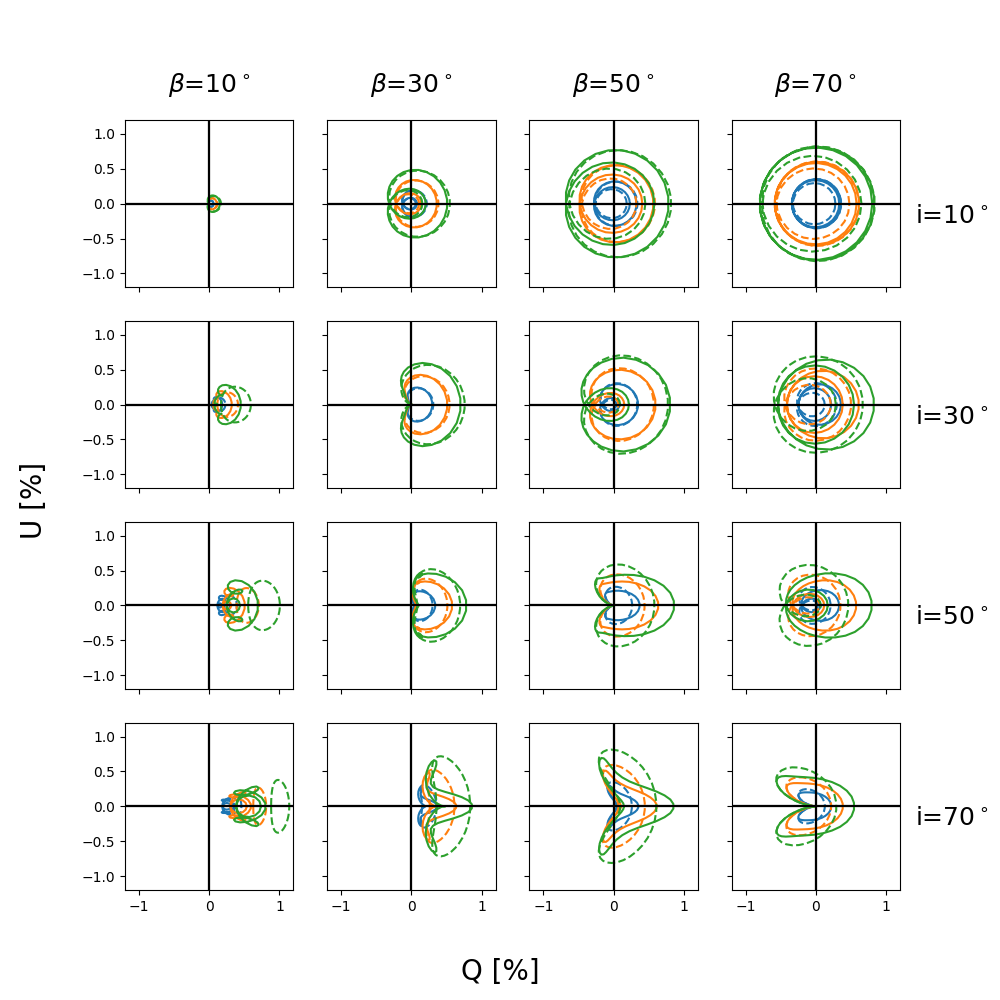}
	\caption{Comparison of Stokes $Q$ and $U$ curves computed in the point light source approximation but reduced by the amplitude ratio in \ref{tab:4} (solid) to curves computed in the finite star regime (dashed). Overplotted are curves of varying $\dot{M}_{B=0}$ while $B_\text{d}$ is left constant. The blue, orange and green lines correspond to values of $\dot{M}=\{1.0,2.0,3.0\}$\Msolyr.}
	\label{fig:11}
	\end{figure*}
	 
	 	 \begin{figure*}
	\includegraphics[width=0.6\linewidth,trim={0 0.25cm 0 1.75cm},clip]{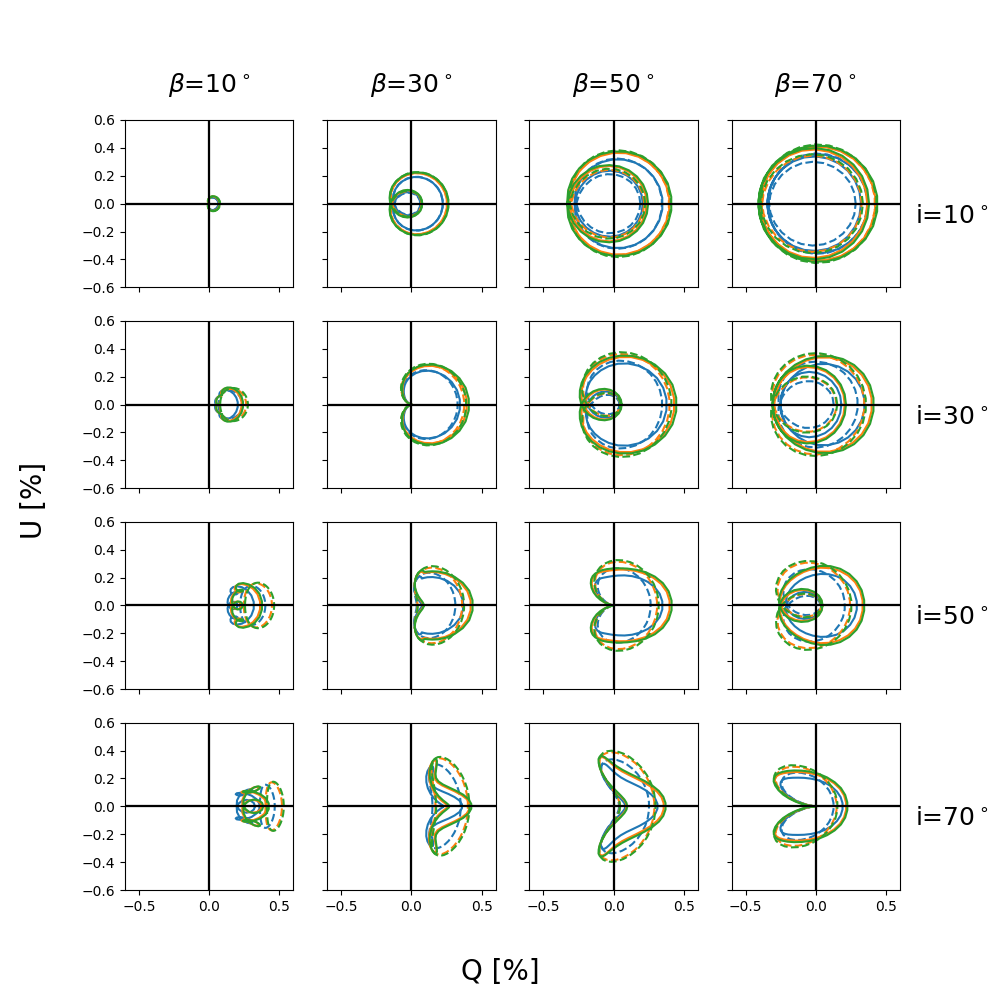}
	\caption{Comparison of Stokes $Q$ and $U$ curves computed in the point light source approximation but reduced by the amplitude ratio in \ref{tab:5} (solid) to curves computed in the finite star regime (dashed). Overplotted are curves of varying $\dot{M}_{B=0}$ while $B_\text{d}$ is left constant. The blue, orange and green lines correspond to values of $\dot{M}_{B=0}=\{1.0,2.0,3.0\}$\Msolyr.}
	\label{fig:12}
	\end{figure*}
    
	\begin{table}
	\centering
	\caption{Amplitude ratios obtained when comparing the Stokes $Q$ and $U$ curves from Fig. \ref{fig:2} to those from Fig. \ref{fig:4}  }
	\label{tab:4}
	\begin{tabular}{cccccc} 
		\hline
		\multicolumn{3}{c}{Input parameters} & \multicolumn{1}{c}{Amplitude Ratio}  \\  
	      $\dot{M}_{B=0}$   & $i$  & $\beta$  & \\
 	  $M_{\odot}$\,yr$^{-1}$ & [deg]& [deg] &          \\
		\hline
         {$1\times10^{-6}$}  & $10$& $10$    &  1.15 \\
                                & $10$& $30$    &  1.10\\
                                & $10$& $50$    &  0.93 \\
                                & $10$& $70$    &  0.76\\
                                & $30$& $10$    &  1.14\\ 
                                & $30$& $30$    &  0.89\\
                                & $30$& $50$    &  0.72\\
                                & $30$& $70$    &  0.61\\
                                & $50$& $10$    &  1.08\\ 
                                & $50$& $30$    &  0.64\\
                                & $50$& $50$    &  0.57\\
                                & $50$& $70$    &  0.56\\
                                & $70$& $10$    &  0.93\\ 
                                & $70$& $30$    &  0.71\\
                                & $70$& $50$    &  0.77\\
                                & $70$& $70$    &  0.62\\
         {$2\times10^{-6}$}  & $10$& $10$    &  1.14\\
                                & $10$& $30$    &  1.10\\
                                & $10$& $50$    &  0.93\\
                                & $10$& $70$    &  0.73\\
                                & $30$& $10$    &  1.2\\ 
                                & $30$& $30$    &  0.90\\
                                & $30$& $50$    &  0.70\\
                                & $30$& $70$    &  0.59\\
                                & $50$& $10$    &  1.10\\ 
                                & $50$& $30$    &  0.63\\
                                & $50$& $50$    &  0.51\\
                                & $50$& $70$    &  0.97\\
                                & $70$& $10$    &  0.70\\ 
                                & $70$& $30$    &  0.75\\
                                & $70$& $50$    &  0.65\\
                                & $70$& $70$    &  0.59\\
          {$3\times10^{-6}$} & $10$& $10$    &  1.20\\
                                & $10$& $30$    &  1.14\\
                                & $10$& $50$    &  0.93 \\
                                & $10$& $70$    &  0.72\\
                                & $30$& $10$    &  1.25 \\ 
                                & $30$& $30$    &  0.90\\
                                & $30$& $50$    &  0.68\\
                                & $30$& $70$    &  0.56 \\
                                & $50$& $10$    &  1.16 \\ 
                                & $50$& $30$    &  0.60 \\
                                & $50$& $50$    &  0.48 \\
                                & $50$& $70$    &  0.49 \\
                                & $70$& $10$    &  0.99 \\ 
                                & $70$& $30$    &  0.69 \\
                                & $70$& $50$    &  0.64 \\
                                & $70$& $70$    &  0.57 \\
		\hline
	\end{tabular}
\end{table}

	\begin{table}
	\centering
	\caption{Amplitude ratios obtained when comparing the Stokes $Q$ and $U$ curves of Fig. \ref{fig:3} to those of Fig. \ref{fig:5} }
	\label{tab:5}
	\begin{tabular}{cccccc} 
		\hline
		\multicolumn{3}{c}{Input parameters} & \multicolumn{1}{c}{Amplitude Ratio}  \\  
	      $ B_\text{d}$   & $i$  & $\beta$ &  \\
  	  $[\text{kG}]$ & [deg]& [deg] &          \\
		\hline
         {$2.5$}  & $10$& $10$    &  0.709 \\
                                & $10$& $30$    &  0.693\\
                                & $10$& $50$    &  0.650 \\
                                & $10$& $70$    &  0.583\\
                                & $30$& $10$    &  0.691\\ 
                                & $30$& $30$    &  0.671\\
                                & $30$& $50$    &  0.638\\
                                & $30$& $70$    &  0.605\\
                                & $50$& $10$    &  0.649\\ 
                                & $50$& $30$    &  0.629\\
                                & $50$& $50$    &  0.625\\
                                & $50$& $70$    &  0.632\\
                                & $70$& $10$    &  0.596\\ 
                                & $70$& $30$    &  0.605\\
                                & $70$& $50$    &  0.701\\
                                & $70$& $70$    &  0.641\\
         {$5.0$}  & $10$& $10$    &  0.719\\
                                & $10$& $30$    &  0.705\\
                                & $10$& $50$    &  0.669\\
                                & $10$& $70$    &  0.611\\
                                & $30$& $10$    &  0.703\\ 
                                & $30$& $30$    &  0.687\\
                                & $30$& $50$    &  0.658\\
                                & $30$& $70$    &  0.628\\
                                & $50$& $10$    &  0.667\\ 
                                & $50$& $30$    &  0.649\\
                                & $50$& $50$    &  0.645\\
                                & $50$& $70$    &  0.653\\
                                & $70$& $10$    &  0.621\\ 
                                & $70$& $30$    &  0.629\\
                                & $70$& $50$    &  0.726\\
                                & $70$& $70$    &  0.677\\
          {$7.5$} & $10$& $10$    &  0.719\\
                                & $10$& $30$    &  0.706\\
                                & $10$& $50$    &  0.673 \\
                                & $10$& $70$    &  0.620\\
                                & $30$& $10$    &  0.705 \\ 
                                & $30$& $30$    &  0.690\\
                                & $30$& $50$    &  0.663\\
                                & $30$& $70$    &  0.635 \\
                                & $50$& $10$    &  0.671 \\ 
                                & $50$& $30$    &  0.655 \\
                                & $50$& $50$    &  0.650 \\
                                & $50$& $70$    &  0.659 \\
                                & $70$& $10$    &  0.628 \\ 
                                & $70$& $30$    &  0.636 \\
                                & $70$& $50$    &  0.716 \\
                                & $70$& $70$    &  0.675 \\
		\hline
	\end{tabular}
\end{table}

	\section{Likelihood distributions}
	
		\begin{figure*}
	\includegraphics[width=12.5cm]{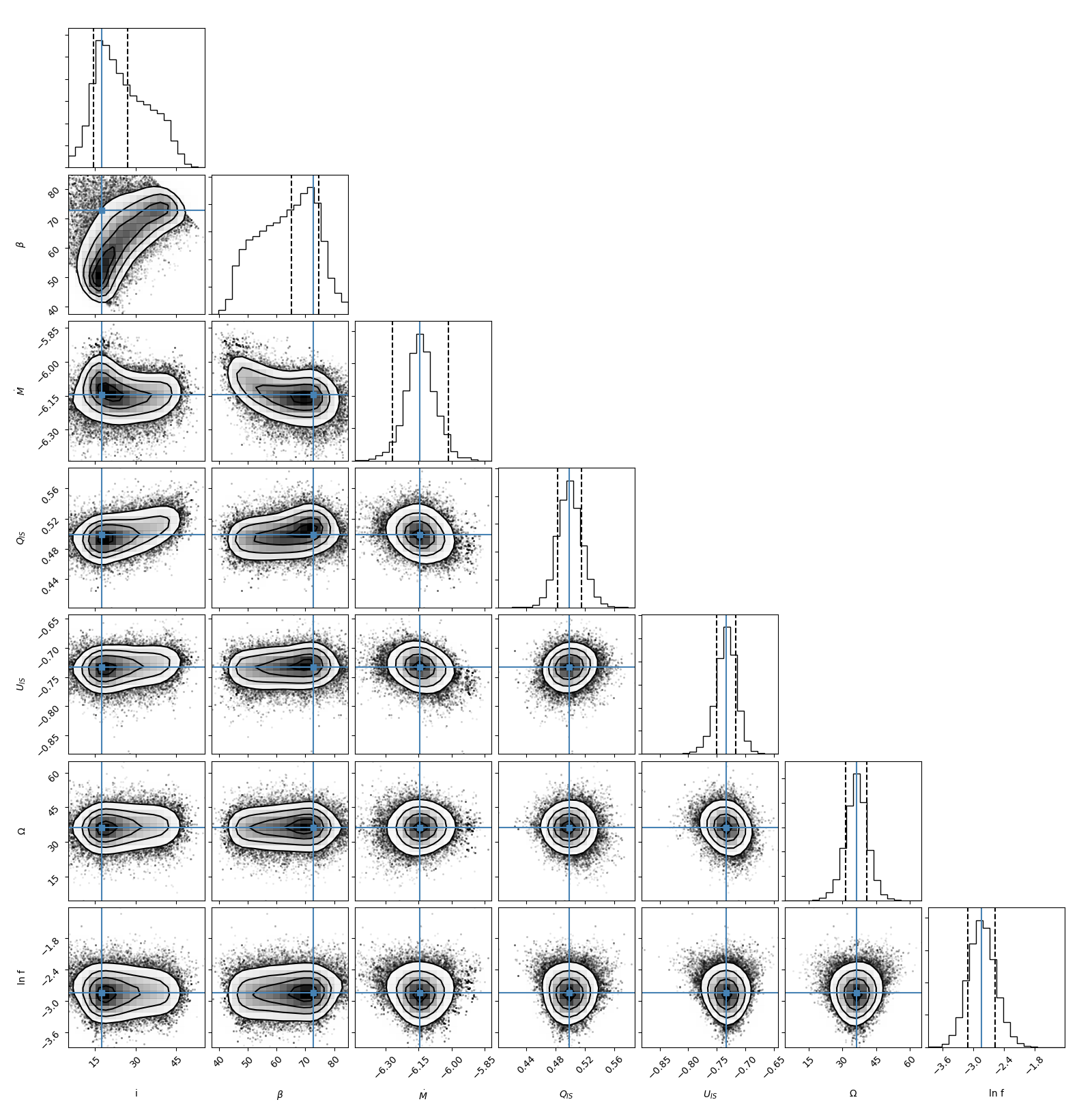}
	\caption{Likelihood distributions for the model parameters of Fig. \ref{fig:6}. Contours are drawn at the 16\%, 50\% and 84\% probability levels and the MCMC-fitted parameters are indicated by the solid (blue) lines.}
	\label{fig:13}
	\end{figure*}

		\begin{figure*}
	\includegraphics[width=12.5cm]{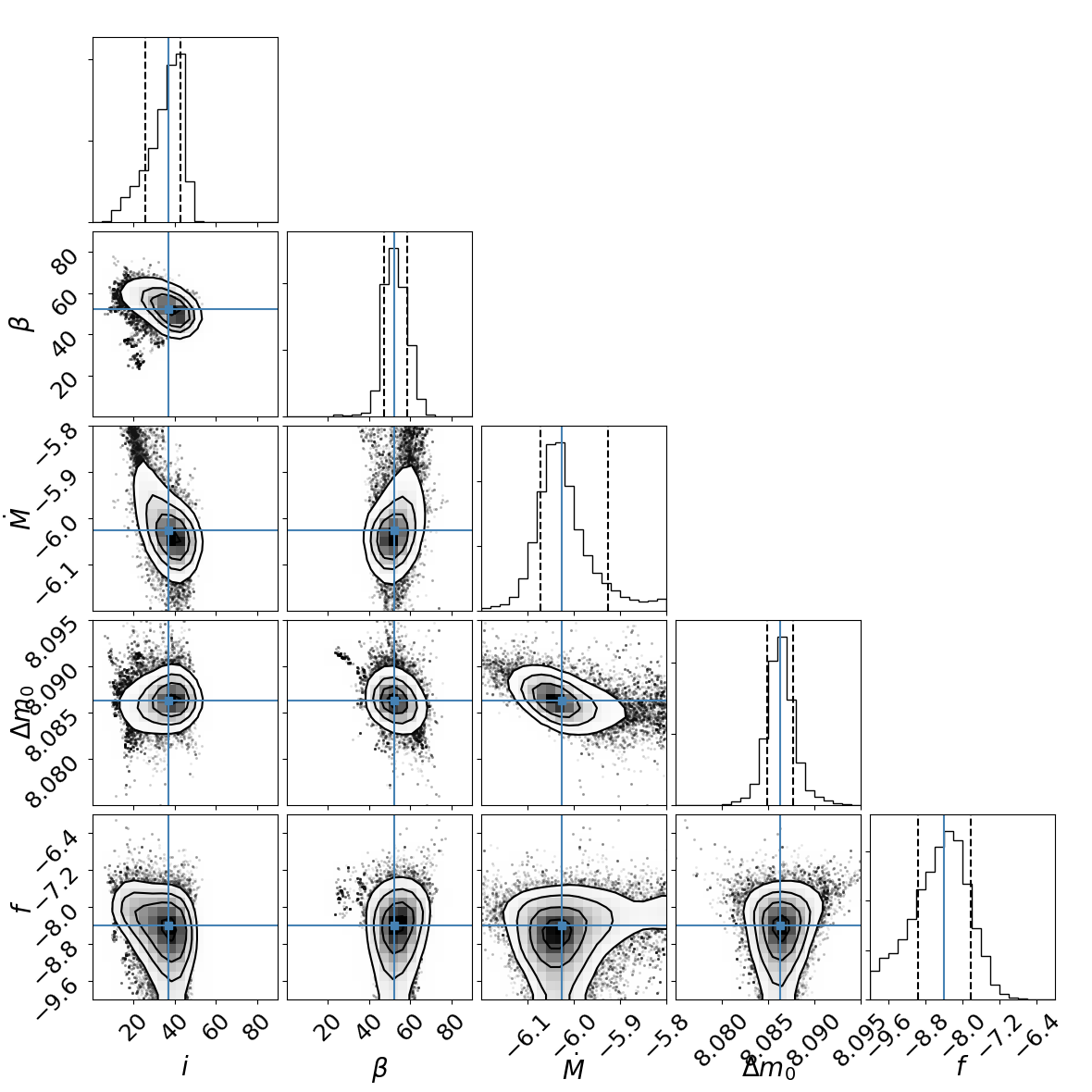}
	\caption{Likelihood distributions for the model parameters of Fig. \ref{fig:7}. Contours are drawn at the 16\%, 50\% and 84\% probability levels and the MCMC-fitted parameters are indicated by the solid (blue) lines.}
	\label{fig:14}
	\end{figure*}	

		\begin{figure*}
	\includegraphics[width=12.5cm]{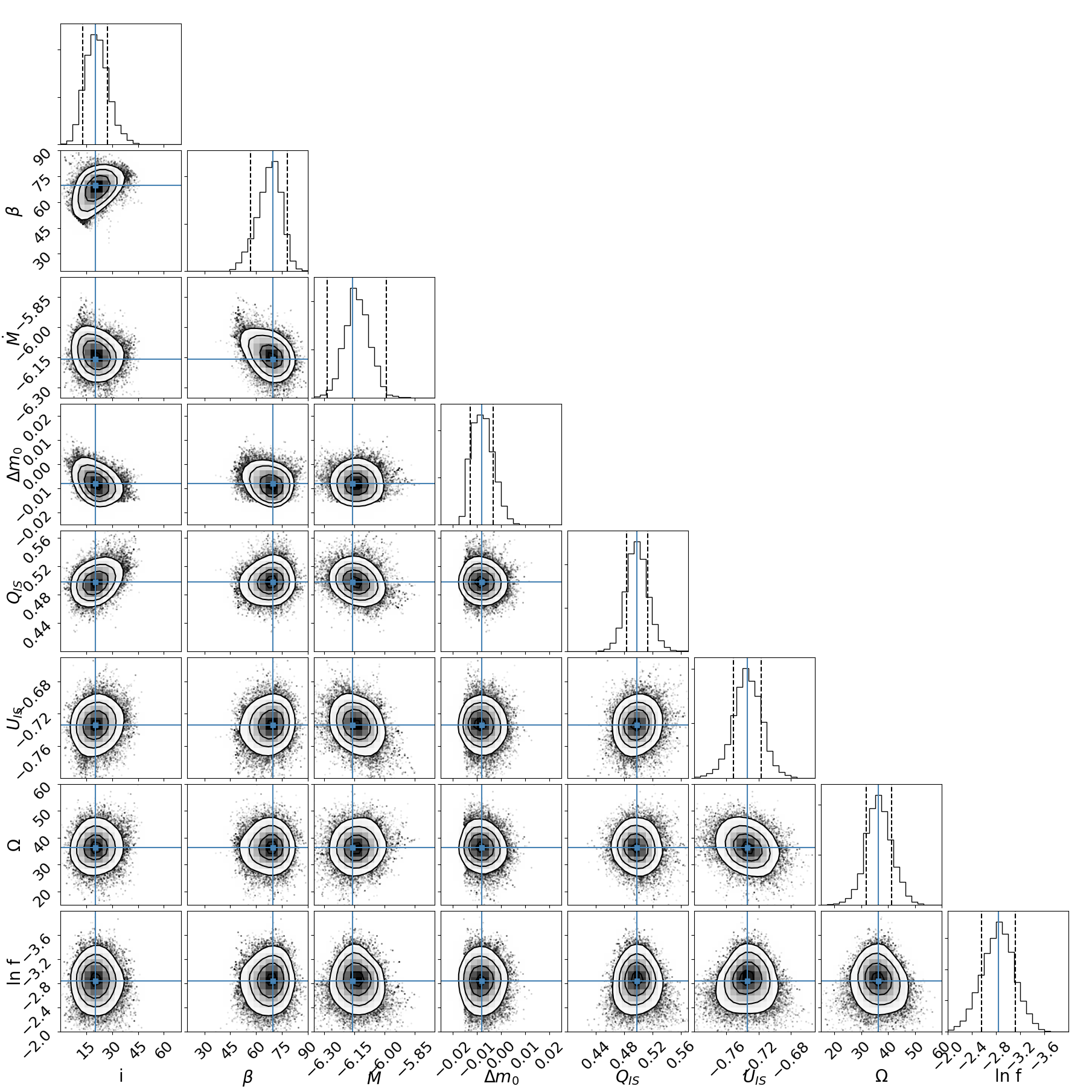}
	\caption{Likelihood distributions for the model parameters of Fig. \ref{fig:8}. Contours are drawn at the 16\%, 50\% and 84\% probability levels and the MCMC-fitted parameters are indicated by the solid (blue) lines.}
	\label{fig:15}
	\end{figure*}

		\begin{figure*}
	\includegraphics[width=12.5cm]{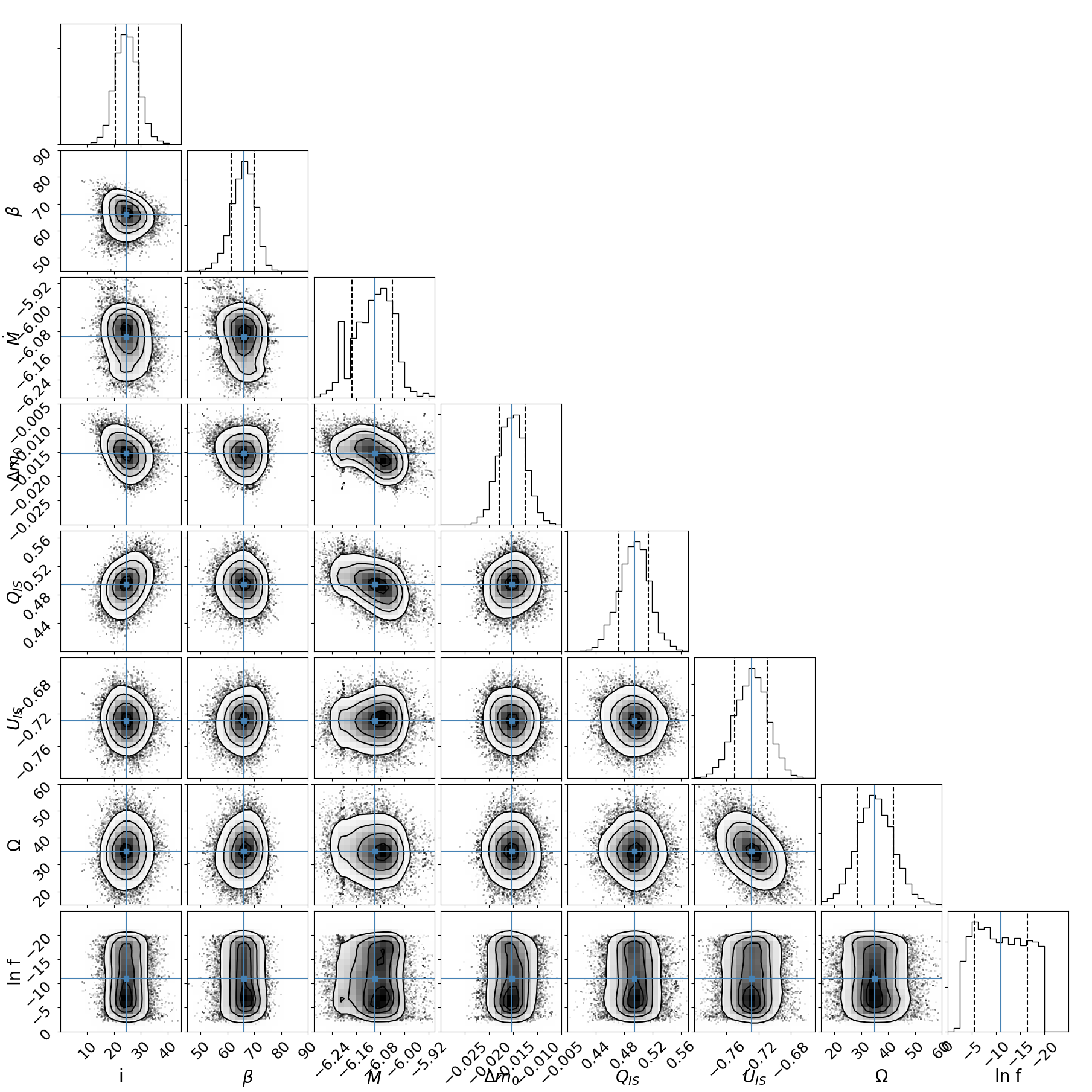}
	\caption{Likelihood distributions for the model parameters of of Fig. \ref{fig:9}. Contours are drawn at the 16\%, 50\% and 84\% probability levels and the MCMC-fitted parameters are indicated by the solid (blue) lines.}
	\label{fig:16}
	\end{figure*}
	
		\begin{figure*}
	\includegraphics[width=12.5cm]{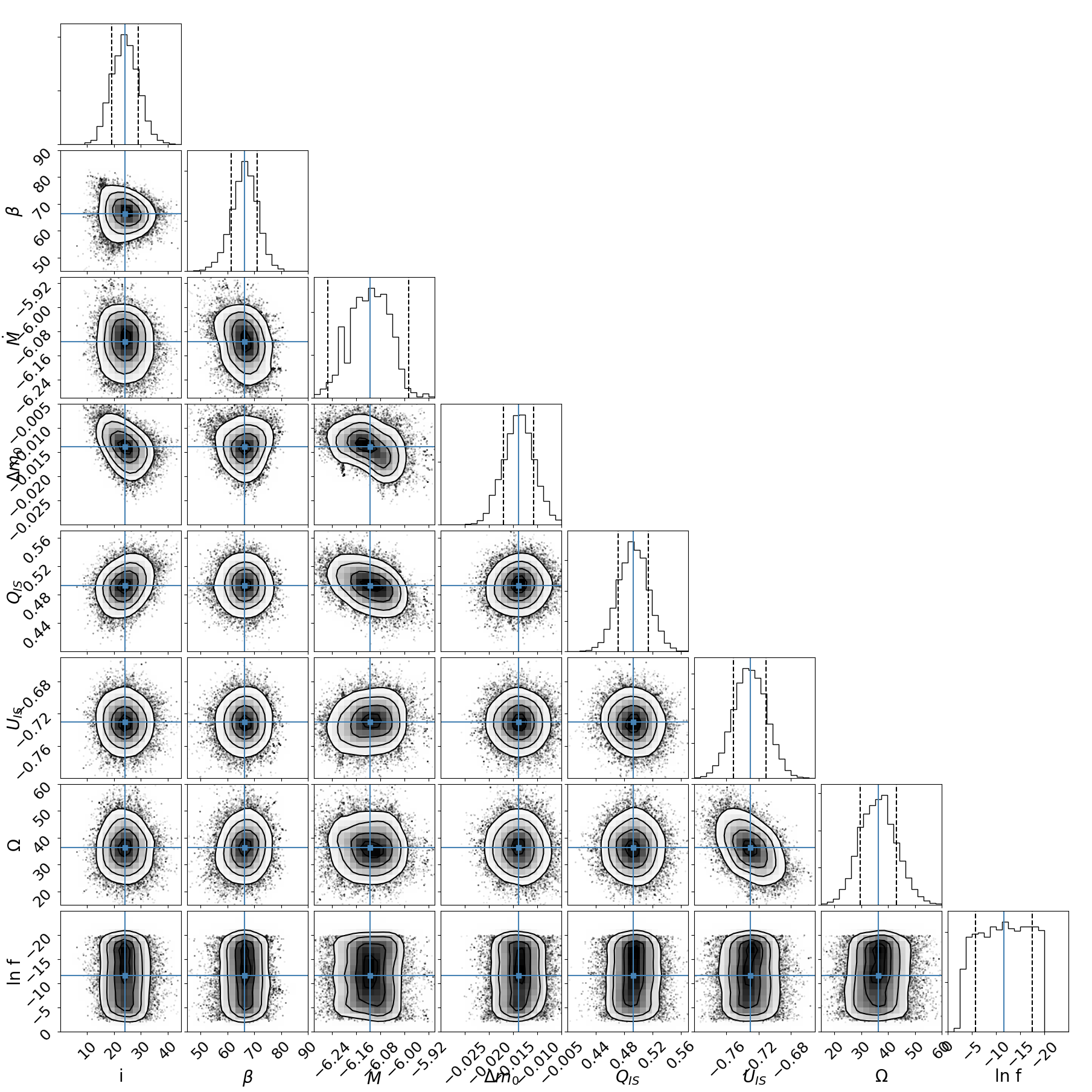}
	\caption{Likelihood distributions for the model parameters of Fig. \ref{fig:10}. Contours are drawn at the 16\%, 50\% and 84\% probability levels and the MCMC-fitted parameters are indicated by the solid (blue) lines.}
	\label{fig:17}
	\end{figure*}

	%%%%%%%%%%%%%%%%%%%%%%%%%%%%%%%%%%%%%%%%%%%%%%%%%%

	% Don't change these lines
	\bsp	% typesetting comment
	\label{lastpage}
	\end{document}